\documentclass{comnet}

\usepackage{amsmath} 
\usepackage{amsthm} 
\usepackage{url}
\usepackage{color}

\usepackage{graphicx}
\usepackage{array}
\newcolumntype{C}[1]{>{\centering\let\newline\\\arraybackslash\hspace{0pt}}m{#1}}

\begin{document}

\title{Evaluating balance on social networks from their simple cycles}

\shorttitle{Evaluating balance on social networks} 
\shortauthorlist{P.-L. Giscard, P. Rochet and R. C. Wilson} 

\author{
\name{Pierre-Louis Giscard$^*$}
\address{University of York, Department of Computer Science, UK\email{$^*$pierre-louis.giscard@york.ac.uk}}
\name{Paul Rochet}
\address{Universit\'{e} de Nantes, Laboratoire de Math\'{e}matiques Jean Leray, France}
\and
\name{Richard C. Wilson}
\address{University of York, Department of Computer Science, UK}}

\maketitle

\begin{abstract}
{Signed networks have long been used to represent social relations of amity (+) and enmity (-) between individuals. Group of individuals who are cyclically connected are said to be balanced
if the number of negative edges in the cycle is even and unbalanced otherwise. 
In its earliest and most natural formulation, the balance of a social network was thus defined from its simple cycles, cycles which do not visit any vertex more than once. Because of the inherent difficulty associated with finding such cycles on very large networks, social balance has since then been studied via other means.  
In this article we present the balance as measured from the simple cycles and primitive orbits of social networks. We specifically provide two measures of balance: the proportion $R_\ell$ of negative simple cycles of length $\ell$ for each $\ell\leq 20$ which generalises the triangle index,  
and a ratio $K_\ell$ which extends the relative signed clustering coefficient introduced by Kunegis.  
To do so, we use a Monte Carlo implementation of a novel exact formula for counting the simple cycles on any weighted directed graph. Our method is free from the double-counting problem affecting previous cycle-based approaches, does not require edge-reciprocity of the underlying network, provides a gray-scale measure of balance for each cycle length separately 
and is sufficiently tractable that it can be implemented on a standard desktop computer.
We observe that social networks exhibit strong inter-edge correlations favouring balanced situations and we determine the corresponding correlation length $\xi$. For longer simple cycles, $R_\ell$ undergoes a sharp transition to values expected from an uncorrelated model. This transition is absent from synthetic random networks, strongly suggesting that it carries a sociological meaning warranting further research.} 
{Social networks; simple cycles; primitive orbits; and Monte-Carlo}
\\
2000 Math Subject Classification: 05C38, 90B10, 91D30,  91G60
\end{abstract}

\section{Introduction}
\subsection{Balance in networks}
Relations of amity and enmity between individuals are well represented by signed networks, where an edge is assigned a positive value if two individuals are acquainted and in good terms, and a negative one if they are instead enemies \cite{Heider1946, Cartwright1956, Harary1959, Norman1972, Harary1980}. Such networks provide a natural setting to study inter-personal relationships and their correlations. 
For example, one could expect that people are friendly towards the friends of their friends, a situation that is said to be  ``balanced".
More generally, on signed networks, a group of individuals who are cyclically connected---i.e. forming a triangle, a square, a pentagon etc.---are said to be balanced if the number of negative edges in the cycle is even. 
Otherwise the cycle is said to be unbalanced. Sociologists have suggested that such negative cycles are the cause of tension and thus, that social
networks should evolve into a state where balanced cycles are largely predominant \cite{Heider1946, Cartwright1956, Antal2006, Aref2015}. The question of whether this holds for real-social networks and if not, by how much this fails to be true, arose from these considerations in the 1940s \cite{Heider1946}. 

Mathematically speaking, this sociological question translates into the following problem: on a signed network $G$, determine for all $\ell$ the percentage of negative simple cycles of length $\ell$.
This problem remains largely unsolved owing to its natural formulation in terms of \textit{simple cycles}---cycles which do not  visit any vertex more than once. 
%
Unfortunately, 
enumerating all the simple cycles of a network \textit{exactly} is computationally intractable. Indeed, this includes counting all the Hamiltonian cycles of the graph, a problem known to be $\#$P-complete \cite{Burgisser1997}. For this reason, we need to seek more efficient methods, which inevitably lead to some approximations. Two strategies are implemented in this work: i) approximate the balance of the network to within any desired accuracy by evaluating the balance on a large sample of subgraphs of the network; or ii) compute the balance exactly from objects which are not simple cycles, but should carry a similar information.\\[-.5em]
 
We successfully implemented the first strategy thanks to a novel exact formula for counting simple cycles on any (weighted directed) graph in conjunction with a Monte Carlo approach. This method is presented in Section \ref{Strat1}. It effectively solves the mathematical problem enunciated earlier since the quality of the obtained approximation is controlled and can be improved at will. 
For the second strategy, we relied on the primitive orbits of the graph, cycles which contain no backtracking steps or tail, and are not the multiple of any other cycle. This is presented in Section \ref{Strat2}. 
The results produced by both approaches on four social networks are discussed and compared in Section \ref{SectionResults}.

\subsection{Notation}
Throughout this article, we consider signed directed networks $G=(\mathcal{V};\mathcal{E})$, of which undirected networks are a special case. The adjacency matrix of $G$ is denoted $\mathbf{A}_G$ or simply $\mathbf{A}$. Each edge of the network is weighted with a value +1 or -1 indicating a
positive or negative interaction. A cycle is positive if the product of its edge
values is positive, and otherwise it is negative. A cycle is simple if it does not visit any vertex more than once. The starting point of a simple cycle is irrelevant but its orientation is retained. 
For example, $v_0v_1v_2v_0$ and $v_1v_2v_0v_1$ represent the same triangle, which is however distinct from $v_0v_2v_1v_0$.
The number of positive and negative simple cycles of length $\ell$ on $G$ are designated by $N_\ell^+$ and $N_\ell^-$, respectively. 

When discussing the balance of a network, we refer to the ratio $R_\ell$ of the number of negatively signed simple cycles of length $\ell$ to the total number of simple cycles of length $\ell$, i.e.
$$
R_\ell := \frac{N_\ell^-}{N_\ell^-+N_\ell^+},\quad \ell\geq 1.
$$
In particular, $R_\ell=0$ when the network is perfectly balanced for length $\ell$, while $R_\ell=1$ indicates a totally unbalanced situation. To facilitate comparisons with existing results, we will also provide the ratio of negative to positive simple cycles
$
\mathcal{U}_\ell := N_\ell^-/N_\ell^+
$
and the relative signed clustering coefficient $K_\ell := (N_\ell^+-N_\ell^-)/(N_\ell^-+N_\ell^+)$ in Appendix~\ref{FullResults}. In the next section, we will see that, for $\ell=3$, the coefficients $R_\ell$ and $K_\ell$ are related to two measures of balance in social networks: the triangle index and the relative signed clustering coefficient (for triads), respectively. 
\subsection{Existing approachs}
The study of balance in social networks has a long history and has been studied via a variety of approaches and we can thus only mention the latest works on the subject here. 
Of fundamental importance was the introduction of grayscale measures of balance, which quantify how balanced any given social network is \cite{Leskovec2010, Kunegis2010, Chiang2014}. 
Many such measures have been proposed, of which we shall briefly review a few important ones here.\\[-.5em] 

The \textit{signed clustering coefficient} and \textit{algebraic conflict} evaluate the balance of a network locally---at the level of triangles---and globally, respectively \cite{Kunegis2010, Kunegis2014}. The former is defined as the number $T^+$ of positive triangles minus the number $T^-$ of negative ones divided by the total number of triplets of vertices connected by at least two edges. In other terms, the signed clustering coefficient takes into account the sign and the tendency to form small clusters (the triangles), putting both considerations on an equal footing. The relative signed clustering coefficient separates the two contributions, being the ratio of $T^+-T^-$ to the total number of triangles, that is exactly the coefficient $K_3$ introduced earlier. The values of $K_\ell$ for $\ell\leq 20$ provided in this study can thus be seen as an extension of the relative signed clustering coefficient to larger clusters of up to 20 individuals.
In general, the relative signed clustering coefficient $K_3$ is exactly accessible via the signed adjacency matrix, see Appendix~\ref{FullResults}.

The algebraic conflict is defined on Harary's basic premise that "in a balanced network, all vertices can be divided into two groups such that  all  positive  edges  connect
vertices  within  the  same  group,  and  all  negative  edges  connect  vertices  of  the
two different groups" \cite{Harary1953}. Kunegis \cite{Kunegis2010}, observed that this condition was fulfilled if and only if the smallest eigenvalue of the graph signed Laplacian is nil. This eigenvalue, called the algebraic conflict, thus carries information about balance and its distance to zero measures the failure of the network to be perfectly balanced.\\[-.5em]

Cycle-based approaches have also been devised, which evaluate various ratios involving the positive and negative cycles of a signed graph. These include the \textit{degree of balance} and its weighted counterpart, arising either from weights attached to the graph edges or from additional length-dependent weights attached to the cycles, see e.g. \cite{Estrada2014}.
The \textit{triangle index} is the fraction of positive triangles on a graph to the total number of triangles. This index is $1-R_3$ and can therefore be seen as a special case of the balance ratios studied in this work. Formally, obtaining the triangle index is a tractable task taking at most $O(N^3)$ time, see Appendix~\ref{FullResults} for a formula giving $R_3$. This computational cost may nonetheless be too high on very large real-life networks. Instead, recent research has shown it could be reliably approximated by computing only the $k$ most dominant eigenvalues of the signed adjacency matrix of the network, thereby reducing the time needed to $O(N^2k)$ \cite{Terzi2011}. Alternatively, the triangle index can be approximated via the Monte Carlo approach presented here.\\[-.5em]

In a 2011 study, Facchetti, Iacono, and Altafini \cite{Facchetti2011} studied the global balance of  social networks using the number of edges whose sign must be changed so that all the simple cycles---and hence all the cycles---be positive. This measure and its close variants are known as the \textit{frustration index} \cite{Iacono2010, Lange2015} and \textit{line index of balance} \cite{Harary1959, Harary1960}. Although these were believed to be computationally intractable to obtain exactly \cite{Facchetti2011}, recent research shows otherwise \cite{Aref2016}, putting these quantities within reach on networks with a few thousands nodes. 
By studying the frustration index, Facchetti \textit{et al.} found that social networks are indeed strongly balanced.
Their conclusions were called into question in a recent work by Estrada and Benzi \cite{Estrada2014} who rather concluded that social networks are not so well balanced using a completely different cycle-based measure of balance, while still other studies employing the frustration index have instead confirmed the conclusion of Facchetti \textit{et al.} \cite{Aref2015}.
As noted earlier, since counting all the simple cycles of a large graph exactly is intractable, one may instead count objects which are not simple but carry a similar information when it comes to balance. 
In this vein, Estrada and Benzi \cite{Estrada2014} proposed the use of
\begin{equation}
D=\operatorname{Tr}\,\exp(\mathbf{A})=\sum_{\ell=0}^\infty \frac{1}{\ell!}\operatorname{Tr}\mathbf{A}^\ell,
\label{Deqn}
\end{equation}
as a method of counting the number of balanced and unbalanced \emph{cycles}, also known as closed walks, in a network. 
They show that, by computing the ratio $K=D/D_+$ with $D_+:=\operatorname{Tr}\,\exp(|\mathbf{A}|)$, called the \textit{weighted degree of balance}, then the ratio of negative to positive cycles can
be obtained as
\begin{equation*}
\mathcal{U}^{\text{walks}}:=\frac{1-K}{1+K}.
\end{equation*}
This is an extremely efficient method which simply requires the evaluation of the eigenvalues
of the adjacency matrix. In practice, this could be expensive on very large networks, but it is sufficient to compute only a few of the largest eigenvalues of  $\mathbf{A}$, as $D$ and $D_+$ are dominated by those \cite{Estrada2014}. \\[-.5em]

\begin{figure}
\begin{center}
\includegraphics[width=0.2\linewidth]{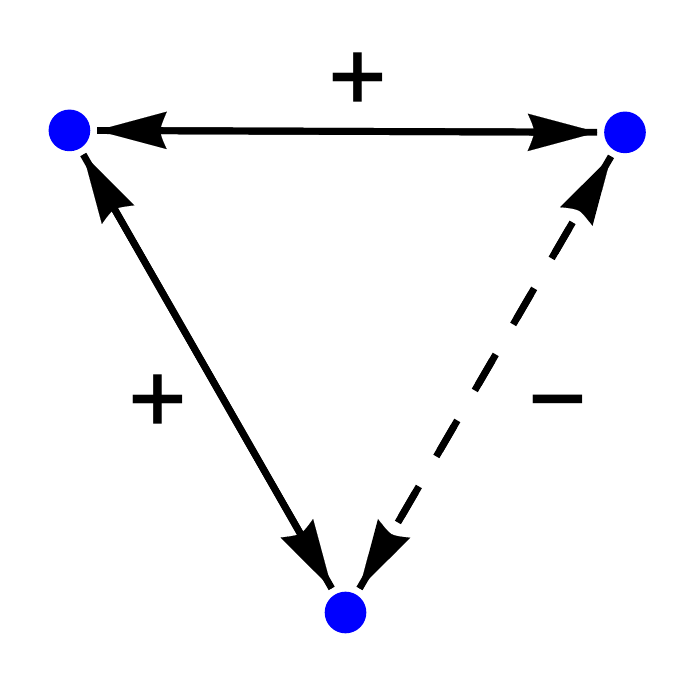}
\end{center}
\vspace{-5mm}
\caption{\label{triad} An unbalanced network of three vertices. The dotted line represents a negative relationship. The exponential unbalance ratio is $\mathcal{U}_{\text{walks}}=0.19$ while in fact $R_2 = 0$ and $R_3 = 1$.}
\end{figure}

Expression (\ref{Deqn}) counts all closed walks (weighted by a factor $1/\ell!$ for a
walk of length $\ell$). Because of this, backtracking steps and multiple cycles are counted. For
example, the non-simple cycle $v_0v_1v_2v_0v_1v_2 v_0=(v_0v_1v_2 v_0)^2$ is part of the sum. 
Such cycles are positive, and so they do not upset the balance of an already balanced network,
but they do have an effect on the (global) balance ratio(s) for an unbalanced network. In other words,
the cycle-sum embodied in (\ref{Deqn}) contains non-simple cycles from the order 2 onwards, effectively mixing the balance ratios $R_\ell$ at all lengths. In addition, the global signature of balance $\mathcal{U}_{\text{walks}}$ further mixes the contributions of the various cycles lengths.
Figure~\ref{triad} illustrates this issue in a triad. Whilst the network is completely unbalanced for triangles, we get $\mathcal{U}^{\text{walks}}=0.19$, a number that is not easy to interpret. 

These difficulties cannot be resolved easily using walks. In an attempt to better account for the length dependency of the balance, we define $D^\ell := \operatorname{Tr}\mathbf{A}^\ell$, $D^\ell_+:=\operatorname{Tr}|\mathbf{A}|^\ell$ and $K^\text{walks}_\ell := D^\ell/D^\ell_+$, $\mathcal{U}^{\text{walks}}_{\ell}:=(1-K^\text{walks}_\ell)/(1+K^\text{walks}_\ell)$ and 
\begin{equation}\label{Rwalks}
R^{\text{walks}}_{\ell}:=\frac{D^\ell_+-D^\ell}{2\,D^\ell_+}.
\end{equation}
These quantities only take walks of length $\ell$ into account when calculating the balance. Yet, since short cycles and their multiples are typically much more abundant than long cycles, the values of $\mathcal{U}^{\text{walks}}_{\ell}$ and $R^{\text{walks}}_{\ell}$ are still largely dominated by the contributions from self-loops, backtracks and triangles. 
Consequently $R^{\text{walks}}_{\ell}$ will be depressed as compared to the true balance ratio $R_\ell$, that is, $R^{\text{walks}}_{\ell}$ overestimates the proportion of balanced cycles. 
We demonstrate this concretely in Section~\ref{SectionResults}, Figures~(\ref{fig:Gama}) and (\ref{fig:Wiki}), where we compare $R^{\text{walks}}_{\ell}$ with $R_{\ell}$ calculated from the simple cycles on two social networks. 
 
While one may empirically argue that long cycles are less relevant than short ones in real social networks \cite{Zajonc1965, Estrada2014}, it seems better to offer as detailed a mathematical analysis as possible before deciding this issue. 
For these reasons, we found it necessary to abandon the use of walks and rather recur either to the simple cycles themselves or to primitive orbits.\\ 

\subsection{Motivating our approach}\label{Motivation}
\subsubsection{On using simple cycles}
Earlier this year, Aref and Wilson \cite{Aref2015} have produced a global study of existing measures of balance. In particular, they recommended the use of the frustration index as a measure of balance and raised several issues affecting cycle-based methods, which we briefly discuss below. 

First, Aref and Wilson make an essential point concerning directed networks: many of the existing cycle-based studies do not take into account the fact that some edges are not reciprocated. It could be that an individual feels in good terms with someone who, in fact, feels negatively towards him. Or it could be that someone has passed a (positive or negative) judgment about someone else, whose attitude towards the first person is not reported in the  data. This makes it difficult to interpret the results of approaches assuming edge-reciprocity, which should thus be, strictly speaking, limited to undirected networks. Second, Aref and Wilson write \cite{Aref2015} ``\textit{cycle-based measures [...] are difficult to compute and do not provide a proper range of values, whether weighted or not.}" Finally, they note that such measures must be improved to avoid or diminish the influence of closed-walks with repeated edges or cycles---which they call cycle double-counting---as these are always balanced.\\[-.5em] 

These observations are central to the present study, which addresses all of them. First, our approach is valid on directed signed networks, i.e. we do not assume nor need reciprocity of the edges\footnote{In other terms, the situations described by Aref and Wilson are exactly taken into account in our analysis. This motivated us to define the balance ratios $R_\ell$, for $\ell\geq 1$ rather than $\ell\geq 3$. And indeed, we will see below that $R_2>0$ in three of the four networks analysed, indicating that the lack of edge-reciprocity does occur in practice}. Remark that an undirected graph is a special case of a directed one, where each directed edge $e_{ij}$ from $i$ to $j$ has a corresponding edge $e_{ji}$ of opposite orientation. It follows that any method for evaluating some indicator of balance that is valid on directed graphs gives rise to a valid approach for undirected graphs.
Furthermore, a simple cycle on an undirected graph corresponds to two cycles with identical signs on the directed version of the same graph. Consequently, the balance ratios $R_\ell$ can always be calculated from the directed form of an undirected graph. 
Second, our approach offers a proper range of values, since $0\leq R_\ell\leq 1$ for all $\ell$, which permits a gray-scale distinction between balanced or unbalanced situations on the one hand and the null hypothesis on the other. In the results, we will see that $R_\ell$ typically varies between 0 and 0.6, exploiting much of its theoretical range. Third, our method is sufficiently tractable that it can be implemented on a standard desktop computer (see below), even on large networks, thanks to a recent breakthrough in the combinatorics of cycles. Finally, our simple-cycle based method is rigorously free from the double-counting problem.

\subsubsection{On studying cycle-lengths separately}
Existing approaches to social balance tend to fall into two categories, both of which provide useful but incomplete information:\\[-.7em]
 
$\blacktriangleright$~Methods based on small structures in social networks, in particular triads, are only sensitive to local interactions in small groups of individuals. For this reason they are unable to detect long range correlations in relationships as well as characterise the dynamics of larger groups of individuals.\\[-1em] 

$\blacktriangleright$~Global measures of balance such as the degree of balance and algebraic conflict, are sensitive to social tensions occurring on all scales (that is group-sizes) but do not yield much information on where those tensions might be, as well as on the groups of individuals typically involved in balanced or unbalanced relations.\\[-.5em] 

The approach recently taken by Facchetti and coworkers \cite{Facchetti2011} is a step towards bridging this divide, as it provides both global and local information on the network balance. Yet, Estrada and Benzi have observed that the frustration index seems to be mostly determined by triads, leaving the influence of larger groups in the final result in doubt. In this work we have taken another route by extending two measures that have been developed for triads, the relative signed clustering coefficient and triangle index, to longer cycles. These longer cycles are sensitive to long range correlations between edge signs, and, by studying each length separately, provide a wealth of information on where social tension occurs by disentangling the influence of the various cycle-lengths on the overall balance.
%
%
%

%

\section{Balance from simple cycles}\label{Strat1}
\label{sec:method}
\subsection{Core combinatorial result}
One possible strategy to estimate the balance ratios $R_\ell$ consists of approximating them from a large sample of subgraphs of the network under study. 
%
The main novelty permitting this straightforward approach in practice is a recently developed mathematical formula for counting simple cycles of any length on weighted directed graphs. 
Rather than sampling the simple cycles directly, the formula allows for a rapid and exact evaluation of the balance ratios on subgraphs of the original network. We show below that this strategy is much better from a computational standpoint than sampling the simple cycles themselves. 

\subsubsection{Formula for counting simple cycles}
Let $P(z)$ be the ordinary generating function of the simple cycles of any weighted directed graph $G$, that is
$$
P(z) := \sum_{c:~\text{simple cycle}} w(c)\, z^{\ell(c)},
$$ 
where $w(c)$ is the weight of $c$, that is the product of the weights of its edges, $z$ is a formal variable and $\ell(c)$ is the length of $c$. 
By exploiting algebraic structures associated with walks on graphs, $P(z)$ can be shown to be 
\begin{equation}
P(z) = \!\!\int\! \frac{1}{z}\!\!\!\!\sum_{H\prec G\atop H\,\text{connected}} \!\!\!\!\operatorname{Tr}\left(\big(z\mathbf{A}_H\big)^{|H|}\big(\mathbf{I}-z\mathbf{A}_H\big)^{|N(H)|}\right)dz.\label{CoreEngine}
\end{equation}
In this expression $H$ is a \emph{connected}\footnote{If $G$ is directed, then the subgraphs $H$ should be \emph{weakly connected} induced subgraphs of $G$. Recall that a digraph is said to be weakly connected if replacing all its directed edges by undirected
edges produces a connected undirected graph.} induced subgraph of $G$, $\mathbf{A}_H$ its adjacency matrix, $|H|$ the number of vertices in $H$ and $|N(H)|$ the number of neighbours of $H$ in $G$. A neighbour of $H$ in $G$ is a vertex $v$ of $G$ which is not in $H$ and such that there exists at least one edge, possibly directed, from $v$ to a vertex of $H$ or from a vertex of $H$ to $v$. The result of Eq.~(\ref{CoreEngine}), as well as further exact formulas for $P(z)$, is presented in \cite{Giscard2016b} and shall not be proven here. 

\subsubsection{Computational cost}
Let $n_H(G)$ be the number of connected induced subgraphs of a graph $G$ and let $n_c(G)$ be the total number of simple cycles on $G$.
The main advantage of Eq.~(\ref{CoreEngine}) is that, if we use a direct algorithm for finding all the connected induced subgraphs of a graph, for example using reverse-search, the time complexity is $O\big(N+|E|\,+N^2 n_H(G)\big)$, with $N$ and $|E|$ the number of vertices and of edges of the graph, respectively \cite{Avis1996,Uehara1999,Elbassioni2015}. In comparison a direct search of the simple cycles themselves must have a time complexity scaling with $n_c(G)$. The best general purpose algorithm for finding such cycles, Johnson's 1975 landmark algorithm \cite{Johnson1975,Mateti1976}, has time complexity  $O\big((N+|E|)(n_c(G)+1)\big)$, and on undirected graphs there exists an algorithm achieving the optimal $O\big(N(n_c(G)+1)\big)$ time complexity \cite{Birmele2013}. In spite of the fact that these algorithms can find the simple cycles, they are also commonly used when counting them is sufficient. In this situation however, Eq.~(\ref{CoreEngine}) with its time complexity scaling with $n_H(G)$ represents a substantial speed-up since in general $n_H(G)\ll n_c(G)$. This is best demonstrated on the complete graph $K_N$ on $N$ vertices---which is the worst case scenario. On $K_N$ we have $n_c(G)\simeq e\times(N-1)!$ where $e\simeq 2.71828$ is the base of the natural logarithm, while $n_H(G)=2^N$ is ``only" exponential\footnote{One should keep in mind that since $P(z)$ determines the existence and number of Hamiltonian cycles on $G$, under the exponential time hypothesis \cite{Impagliazzo2001} this exponential cost is, in principle, the best possible.} Furthermore, in practice, the computational cost is much smaller as real networks are typically sparse.

Besides Johnson's algorithm, the most efficient \emph{general purpose}\footnote{In particular, the remarkable algorithm of Alon, Yuster and Zwick \cite{Alon1997}, which is extremely efficient for counting simple cycles, is not a general purpose one. Rather, it is limited to undirected graphs and cycle lengths of at most 7.} algorithm for counting simple cycles is that of Bax and Franklin \cite{bax1996finite}. This algorithm evaluates a formula counting the simple cycles that involves a sum over all induced subgraphs of the graph, as opposed to only the \textit{connected} ones appearing in Eq.~(\ref{CoreEngine}). On sparse graphs, there are far more
induced subgraphs than connected induced subgraphs and evaluating Eq.~(\ref{CoreEngine}) yields a significant speed-up as compared to Bax and Franklin's algorithm.\\

Most importantly for applications, Eq.~(\ref{CoreEngine}) is well suited to truncations: only those connected induced subgraphs $H$ of $G$ for which $|H|\leq\ell\leq |H|+|N(H)|$ can possibly contribute to the coefficient of $z^\ell$ in $P(z)$. This means that if one is interested in the first $\ell$ terms of $P(z)$---that is in the simple cycles of length up to $\ell$---it suffices to consider those connected induced subgraphs of $G$ with $|H|\leq \ell$. Using reverse search, this has time complexity $O(N+|E|+\ell^2 n_{H:\,|H|\leq \ell}(G))\leq O(N^2+\ell^2 n_{H:\,|H|\leq \ell}(G))$ \cite{Uehara1999,Elbassioni2015}. Since furthermore only the small adjacency matrices $\mathbf{A}_H$ enter Eq.~(\ref{CoreEngine}), each term of the equation costs $O(|H|^3)\leq O(\ell^3)$ to evaluate. 
Thus, getting the first $\ell$ terms of $P(z)$ from Eq.~(\ref{CoreEngine}) has time complexity bounded by $O(\ell^3N^{\ell})$ in the worst case scenario of the complete graph, and \emph{far less} on sparse graphs. Instead, the computational cost on sparse graphs can be bounded as follows: let $\Delta$ be the maximum vertex-degree on the network. Then the number of connected induced subgraphs  on at most $\ell$ vertices is bounded by $O\big(N \Delta^{\ell}/((\Delta-1)\ell^2)\big)$ \cite{Uehara1999}. In addition, the time complexity of reverse search in this case is $O(N(\Delta+1)+N \Delta^{\ell})$ \cite{Avis1996,Uehara1999,Elbassioni2015}. Consequently Eq.~(\ref{CoreEngine})  can be shown to produce $P(z)$ exactly up to order $\ell$ in $O\big(N\Delta^\ell/\ell\big)$ time.\footnote{The properties of Eq.~(\ref{CoreEngine}) contrast it with other analytical formulas for counting simple cycles of small lengths \cite{Harary1971,Movarraei2016,Alon1997}. First, these formulas work only on undirected graphs. Second, these formulas comprise a large number of terms, e.g. 160 terms for simple cycles of length 10, all of which involve the adjacency matrix of the full graph. 
} For a more detailed analysis of the algorithmic implementation of Eq.~(\ref{CoreEngine}) together with comparison with other algorithms for the same task, see  \cite{Giscard2016c}.\\

To give concrete examples, with an Intel Core i7-4790 CPU @ 3.60 GHz desktop computer, evaluating Eq.~(\ref{CoreEngine}) on the complete graph on $15$ vertices took on average $\sim\!0.7\,$sec, yielding $255,323,504,932\simeq2.5\times10^{11}$ for the total number of simple cycles, which we  verify analytically to be exact. This went up to $\sim\!25\,$sec for the complete graph on $20$ vertices where  $349,096,664,728,623,336\simeq 3.5\times 10^{17}$ simple cycles were counted, a number that is, once again, exact. 
In both cases about half of the computer time was spent looking for the connected induced subgraphs and the other half implementing Eq.~(\ref{CoreEngine}).
For the real-world networks analysed below, a randomly chosen induced subgraph on 30 vertices is typically analysed in $0.001-0.3\,$seconds on the same computer, depending on its sparsity.


\subsection{Monte Carlo implementation}
When the size of the network to study is large---what ``large" means here strongly depends on the graph sparsity---an exact calculation of the desired terms of $P(z)$ from Eq.~(\ref{CoreEngine}) becomes intractable and the core combinatorial result must be supplemented by a Monte Carlo approach.\\[-.5em]

The reader may have noticed upon close inspection of Eq.~(\ref{CoreEngine}) that $P(z)$ results from subtle cancellations between the contributions of the various connected induced subgraphs $H$ of $G$. For this reason,  Eq.~(\ref{CoreEngine}) is not directly amenable to a Monte Carlo method which would consist of randomly selecting a sample of connected induced subgraphs $H$ of the whole network $G$ and estimating $P(z)$ from this sample.  Eq.~(\ref{CoreEngine}) can however be evaluated very quickly on graphs of ``reasonable" size---once again this depends on the sparsity of the underlying graph and the available computational resources.\\[-.5em] 

Our strategy is therefore to sample $N$ induced subgraphs of the network under study and to calculate the balance ratios $R_\ell$ \emph{exactly} up to the desired length $\ell$ on each of these samples via Eq.~(\ref{CoreEngine}). 
The average value of all the $R_\ell$ then converges to that of the whole network as $N$ grows. The quality of this approximation is appraised by repeating the whole procedure $N'$ times and extracting the standard deviation on the averaged $R_\ell$. If this deviation is too large, the number $N$ of samples is increased and the standard deviation is reevaluated. Once the deviation is below the desired accuracy, the method is deemed to have converged. We also systematically tested the method against bias by comparing it with exact results whenever available, see Appendix~\ref{AlgoResultChecks}.\\[-.5em]

Firstly, we decided to limit our study to simple cycles of length up to 20. Longer cycles are both rare and largely irrelevant since, as will see, at such lengths $\ell>20$ all the social networks follow the null hypothesis. There remains three important parameters controlling the Monte Carlo approach which we can choose using a qualitative analysis:
\begin{itemize} 
\item[1)] \underline{Number of samples $N$:} it is the total number of subgraphs on which the simple cycles are counted. 
The larger this number, the more simple cycles are found and the smaller the variance $\text{var}(R_\ell)$ of the $R_\ell$ ratios, that is the better the results. Following well established principles, we expect $\text{var}(R_\ell)\propto 1/\sqrt{|\text{Cycle}_\ell|}$, $|\text{Cycle}_\ell|$ being the total number of simple cycles of length $\ell$ found in all samples. 

\item[2)] \underline{Size $N_{\text{vertices}}$ of each sample:} it is the number of vertices in each sampled subgraph. Since the longest simple cycle on a graph on $N_{\text{vertices}}$ has length $\ell\leq N_{\text{vertices}}$, we must have $N_{\text{vertices}}\geq 20$. Letting $f_\ell(N_{\text{vertices}})$ be the average number of simple cycles of length $\ell$ found on all sampled subgraphs, we have $|\text{Cycle}_\ell| = N_{\text{sample}} f_\ell(N_{\text{vertices}})$. In addition, we can expect $f_\ell$ to be monotonous increasing function of $N_{\text{vertices}}$, in particular $f_\ell(N_{\text{vertices}})$ is at most $O(N_{\text{vertices}}^\ell)$ on complete graphs.

\item[3)]\underline{Time $\tau_\ell(N_{\text{vertices}})$ spent per sample:} we have shown in the preceding section that this time obeys  
$
\tau_\ell(N_{\text{vertices}}) = O\left(N_{\text{vertices}} \Delta^{\ell}/\ell\right),
$
where $\Delta$ is the average maximum degree of any vertex in the sampled subgraph. On average, $\Delta\leq N_{\text{vertices}}$ is expected to be a monotonous increasing function of $N_{\text{vertices}}$.
\end{itemize}

Taking these three quantities into account, we end up with the following trade-off between accuracy of the results and time taken to reach them
$$
\text{var}(R_\ell) \propto \frac{1}{\sqrt{N \,f_\ell(N_{\text{vertices}})}}\iff \text{Total time}= O\left(N \frac{N_{\text{vertices}}^{\ell+1}}{\ell}\right),
$$
from which we deduce that $N$ should be large while $N_{\text{vertices}}$ should be  as small as possible. Indeed, the time taken to count simple cycles of length $\ell$ can scale as badly as $N_{\text{vertices}}^{\ell+1}$ while the variance cannot diminish faster than $\sqrt{f_\ell(N_{\text{vertices}})}\leq N_{\text{vertices}}^{\ell/2}$. In consequence increasing $N$ is a better way to improve the accuracy of the results than increasing the size of each sampled subgraph, since the overall computation time scales only linearly with $N$.\\

Empirically, we confirmed the qualitative analysis outlined above as we found that
a large number $N$ of samples of size $N_{\text{vertices}}\sim 20-30$ vertices offered the best trade-off between accuracy and computation time. Note, in practice 
$\tau_{20}(N_{\text{vertices}})$ is found to be between a millisecond and a tenth of a second time.

~\\ 

\section{Balance from primitive orbits}\label{Strat2}

\subsection{Background}
A \emph{primitive orbit} \footnote{Unfortunately, primitive orbits are also known as ``prime cycles", which is, strictly speaking, a misnomer. Indeed, primitive orbits do not satisfy the fundamental definition of a prime element, namely $p$ is prime if and only if $p|a.b\iff p|a$ or $p|b$ for all $a,\,b$. Ironically, the only objects obeying this definition on a graph are the simple cycles, see \cite{Giscard2016}.} on a network is a cycle which contains no backtracking steps or tail, and
is not a recurrence of any other cycle, e.g. $(v_0v_1v_2 v_0)^2$ \cite{Terras2011}. It is important to note that this is not the same as
a simple cycle---for example if $c_1$ and $c_2$ are simple cycles sharing an edge,
then $c_1c_2$ is a primitive orbit. However, primitive orbits are identical to simple cycles up
to order five \footnote{This is true if and only if the graph has no self-loops, i.e. length 1 cycles. In the presence of such loops, we can simply remove them by replacing $\mathbf{A}$ by $\mathbf{A}-\text{Diag}(\mathbf{A})$ in the calculations of $R_{\ell\geq 3}$.}. The (inverse) Ihara zeta function of a graph is given by \cite{Ihara1966}
\begin{equation}
\zeta^{-1}_{|G|}(z)=\prod_{c\in[C]} \left(1-z^{\ell(c)}\right),
\end{equation}
where $\ell(c)$ is the length of the primitive orbit and $|G|$ denotes the unsigned version of $G$. The notation $[C]$ designates the set of equivalence classes of primitive orbits, i.e. the set of primitive orbits where all starting points on the same cycle are considered equivalent. The zeta function $\zeta_{|G|}^{-1}(z)$ can be
expanded into terms representing each primitive orbit; to first order the expansion
is $\zeta_{|G|}^{-1}(z)=1-\sum_{c\in[C]} z^{\ell(c)}+ O(z^6)$. Therefore, up to order 5, the zeta function corresponds to a sum over simple cycles, and non-simple cycle contamination only occurs at order 6 and higher. 
Because of this, the Ihara zeta function will provide a more precise measure of balance than a walk-based one.

To approximate the balance ratios $R_\ell$ using primitive orbits, we begin by introducing a modified
version of the Ihara zeta function for signed networks
\begin{equation*}
\zeta^{-1}_G(z)=\prod_{c\in[C]} \left(1-s(c)z^{\ell(c)}\right),
\end{equation*}
where $s(c)$ is the sign of the primitive orbit $c$. We now need an efficient way of
evaluating the zeta function. This can be achieved using its determinant form
\begin{equation*}
\zeta^{-1}_G(z)=\det\left(\mathbf{I}-z\mathbf{T}\right),
\end{equation*}
where $\mathbf{T}$ is the Hashimoto matrix of the network, also known as edge adjacency matrix \cite{Hashimoto1989}. 
Let us recall the standard definition of $\mathbf{T}$ here.
We designate by $e_i$, $1\leq i\leq |E|$, an edge of the graph equipped with a randomly chosen orientation, and by $e_i^{-1}$ the edge $e_i$ with the opposite orientation. 
Then $\mathbf{T}$ is the $2|E|\times 2|E|$ matrix defined as \cite{Terras2011}:
$$\mathbf{T}_{ij}:= \begin{cases}1,&\text{if and only if the end vertex of edge $e_i$ is the start vertex of edge $e_j$ and $e_j\neq e_i^{-1}$}\\
0,&\text{otherwise.}
\end{cases}
$$
In other words, $\mathbf{T}$ is the
adjacency matrix of the oriented line graph (OLG), that is
its vertices correspond to the directed edges of the original graph. 
In the OLG, vertices are connected if there is an allowed two-step walk along the corresponding edges in the original graph and backtracking steps are not allowed
(so $ab,bc$ is allowed if $ab$ and $bc$ are edges, but $ab,ba$ is not allowed).

To incorporate the edge signs into the matrix, we use forward sign assignment. Since
all terms in the zeta function are cycles, we can uniquely place the sign of edge
$ab$ into any edge in $\mathbf{T}$ which begins from $ab$, and the sign of the
cycle in $\mathbf{T}$ will be the same as the sign of the original cycle.


\subsection{Computation}
In principle, the number of primitive orbits of any length is easily determined from traces of powers of $\mathbf{T}$. We have the following result, which we prove in Appendix \ref{Proof}.
\begin{proposition}\label{PropNT}
Let $G$ be a signed directed graph, $\mathbf{T}$ its Hashimoto adjacency matrix and $N_{\text{ob};\,\ell}^+$ and $N_{\text{ob};\,\ell}^-$ be the number of positive and negative primitive orbits of length $\ell$ on $G$, respectively.  Then
\begin{align*}
N_{\text{ob};\,\ell}^+-N_{\text{ob};\,\ell}^-&= \frac{1}{\ell}\sum_{d|\ell}\mu(\ell/d) \operatorname{Tr}\mathbf{T}^d,
\end{align*}
where $\mu(.)$ is the number-theoretic M\"{o}bius function. A similar result holds for $N_{\text{ob};\,\ell}^++N_{\text{ob};\,\ell}^-$ upon replacing $\mathbf{T}$ by $|\mathbf{T}|$.
\end{proposition}

This formula is particularly revealing as to the connection between the Hashimoto matrix and the primitive orbits. Traces of powers of $\mathbf{T}$ count
all cycles in the OLG, i.e. the backtrackless closed walks, including the so-called power orbits, e.g. $(c_1)^2$ and $(c_1 c_2)^2$, which are not primitive.
The set of cycles of length $\ell$ contains such power orbits if and only if there are orbits whose length is a divisor of $\ell$. These divisors are removed via a M\"{o}bius inversion in the above sum.
For moderately sized graphs, it is straightforward to compute $\mathbf{T}$ and its spectrum, and so to compute the number of primitive
orbits of any length. However, since $\mathbf{T}$ is the adjacency of the OLG, its size is equal to the number of
edges in the original network. In practice, it can therefore become difficult to compute the eigenvalues
of this matrix. Since $ \operatorname{Tr}\mathbf{T}= \operatorname{Tr}\mathbf{T}^2=0$, we need only concern ourselves with
third and higher powers of the eigenvalues. The spectrum can therefore be effectively truncated, considering only the largest
magnitude eigenvalues, thereby reducing the computational burden. Nevertheless, this can be computationally demanding for large networks.

For undirected unsigned graphs, Stark and Terras \cite{Stark1996} provide a way to compute the number of all orbits (i.e. $ \operatorname{Tr}\mathbf{T}^\ell$), which we adapted to count the number of positive and negative primitive orbits in a signed but \emph{undirected} network. Let $\mathbf{D}$ be the diagonal degree matrix of $|G|$ and
let $\mathbf{Q}=\mathbf{D}-\mathbf{I}$. Further let $\mathbf{A}^+$ be the adjacency
of only the positive edges in $G$ and similarly $\mathbf{A}^-$ be the adjacency of
only the negative edges in $G$ (coded as -1). Then the following iteration counts
the number of positive and negative backtrackless closed walks (orbits) between any two vertices, denoted $W_{\text{ob};\,\ell}^+$ and $W_{\text{ob};\,\ell}^-$, 
\begin{eqnarray*}
\mathbf{A}^+_2&=&\mathbf{A}^+\mathbf{A}^++\mathbf{A}^-\mathbf{A}^--(\mathbf{Q}+\mathbf{I}), \nonumber \\
\mathbf{A}^-_2&=&\mathbf{A}^-\mathbf{A}^++\mathbf{A}^+\mathbf{A}^-, \nonumber \\
\mathbf{A}^+_\ell&=&\mathbf{A}^+_{\ell-1}\mathbf{A}^++\mathbf{A}^-_{\ell-1}\mathbf{A}^--\mathbf{A}^+_{\ell-2}\mathbf{Q}, \nonumber \\
\mathbf{A}^-_\ell&=&\mathbf{A}^-_{\ell-1}\mathbf{A}^++\mathbf{A}^+_{\ell-1}\mathbf{A}^--\mathbf{A}^-_{\ell-2}\mathbf{Q}, \nonumber \\
W_{\text{ob};\,\ell}^+&=&\operatorname{Tr}\left(\mathbf{A}^+_\ell-(\mathbf{Q}-\mathbf{I})\sum_{j=1}^{(\ell-1)/2}\mathbf{A}^+_{\ell-2j}\right), \nonumber \\
W_{\text{ob};\,\ell}^-&=&\operatorname{Tr}\left(\mathbf{A}^-_\ell-(\mathbf{Q}-\mathbf{I})\sum_{j=1}^{(\ell-1)/2}\mathbf{A}^-_{\ell-2j}\right).
\end{eqnarray*}
Since $W_{\text{ob};\,\ell}^+-W_{\text{ob};\,\ell}^-=\operatorname{Tr}{\mathbf{T}}^\ell$ and $W_{\text{ob};\,\ell}^++W_{\text{ob};\,\ell}^-=\operatorname{Tr}{|\mathbf{T}|}^\ell$, using these results in conjunction with Proposition~\ref{PropNT} counts the primitive orbits. This method is very efficient thanks to its use of $\mathbf{A}$ rather than $\mathbf{T}$, but remains limited to undirected networks.\\ 

Armed with the number of positive and negative primitive orbits of length $\ell$, we compute the primitive-orbits-based ratios 
$ R^{\text{ob}}_{\ell\geq 3}:= N_{\text{ob};\,\ell}^-/(N_{\text{ob};\,\ell}^-+N_{\text{ob};\,\ell}^+)$. We will see that these provide better approximations of the true balance ratios $R_\ell$ than walk-based ones $R^{\text{walks}}_{\ell}$.

~\\

\section{Results}\label{SectionResults}
In this section we present the evolution of the balance ratios $R_\ell$ with $\ell$ on several social networks. As detailed in \S\ref{Motivation}, we aim, on each network, at determining the evolution of $R_\ell$ with $\ell$. We will see that on all the large social networks studied here, $R_\ell$ exhibits a sharp transition around $\ell\sim 10-12$, which is absent from synthetic networks. We discuss the relation between this observation and past studies in \S\ref{CompareInterprete}, concluding with a possible structural interpretation for it. Tables with full numerical results for $R_\ell$, $K_\ell$ and $\mathcal{U}_\ell$ are included in Appendix~\ref{FullResults}.

\subsection{Data sets}
Following the precedent studies by Facchetti \textit{et al.} \cite{Facchetti2011}, Estrada and Benzi \cite{Estrada2014} and Aref and Wilson \cite{Aref2015}, we have analysed four social networks: i) Gahuku-Gama with 16 vertices \cite{Gama}; ii) WikiElections with 8297 vertices \cite{StanfordWiki}; iii) Slashdot with 82,144 vertices \cite{StanfordSlashdot}; and iv) Epinions with 131,828 vertices \cite{StanfordEpinions}. Note that among these, only the Gahuku-Gama network is undirected. 
%

\subsection{Null-hypothesis}
In order to meaningfully determine if social networks are balanced, we compare our results to the balance 
 that would be obtained on a graph with the same proportion $p$ of negative directed edges than the real network under study, but where the sign of any directed edge is negative with probability $p$. In particular,  in the null-hypothesis model, the signs of any two directed edges are \textit{independent} random variables. Then the probability that a simple cycle $c$ of length $\ell$ be negative is
\begin{equation}
\text{Prob}(c~\text{negative}) = \sum_{i=0}^{\lceil\ell/2\rceil-1}\binom{\ell}{i} p^{2i+1}(1-p)^{\ell-2i-1}.\label{Rnull}
\end{equation}
Supposing for simplicity that the signs of any two simple cycles are independent random variables then the probability distribution for $N^-_\ell/(N^-_\ell+N^+_\ell)$ in the null-hypothesis is a binomial law with expectation value $R_\ell^{\text{null}}$ given by Eq.~(\ref{Rnull}). Consequently, in this simple model the null-hypothesis is compatible up to a near 95$\%$ confidence level with any value of $R_\ell$ within the 2$\sigma$ interval
\begin{equation}\label{ErrorNull}
R_\ell^{\text{null}}\pm 2 \frac{\sqrt{R_\ell^{\text{null}}(1-R_\ell^{\text{null}})}}{\sqrt{N^-_\ell+N^+_\ell}}.
\end{equation}

The assumption that the signs of any two simple cycles are independent random variables is not true on real social networks. Calculating the null-hypothesis without this assumption is very difficult in practice however. Indeed, a more accurate null model is given by evaluating the average balance ratios of all lengths over all random shufflings of the edges-signs from the social network under study.
We implemented this more accurate model on the WikiElections network and found it to yield null balance ratios that are up to $9\%$ lower than the values predicted by Eq.~(\ref{Rnull}) when $\ell\lesssim 10$, while differences diminish for longer simple cycles. Yet, all the conclusions that can be drawn from comparing the simple null model Eqs.~(\ref{Rnull}, \ref{ErrorNull}) with the computed balance ratios are unchanged, since the relative positions of the two are unaltered by the more accurate model.\\



\subsection{Gahuku-Gama network}\label{GamaSection}
\begin{figure}[t!]
\begin{center}
\includegraphics[width=.9 \textwidth]{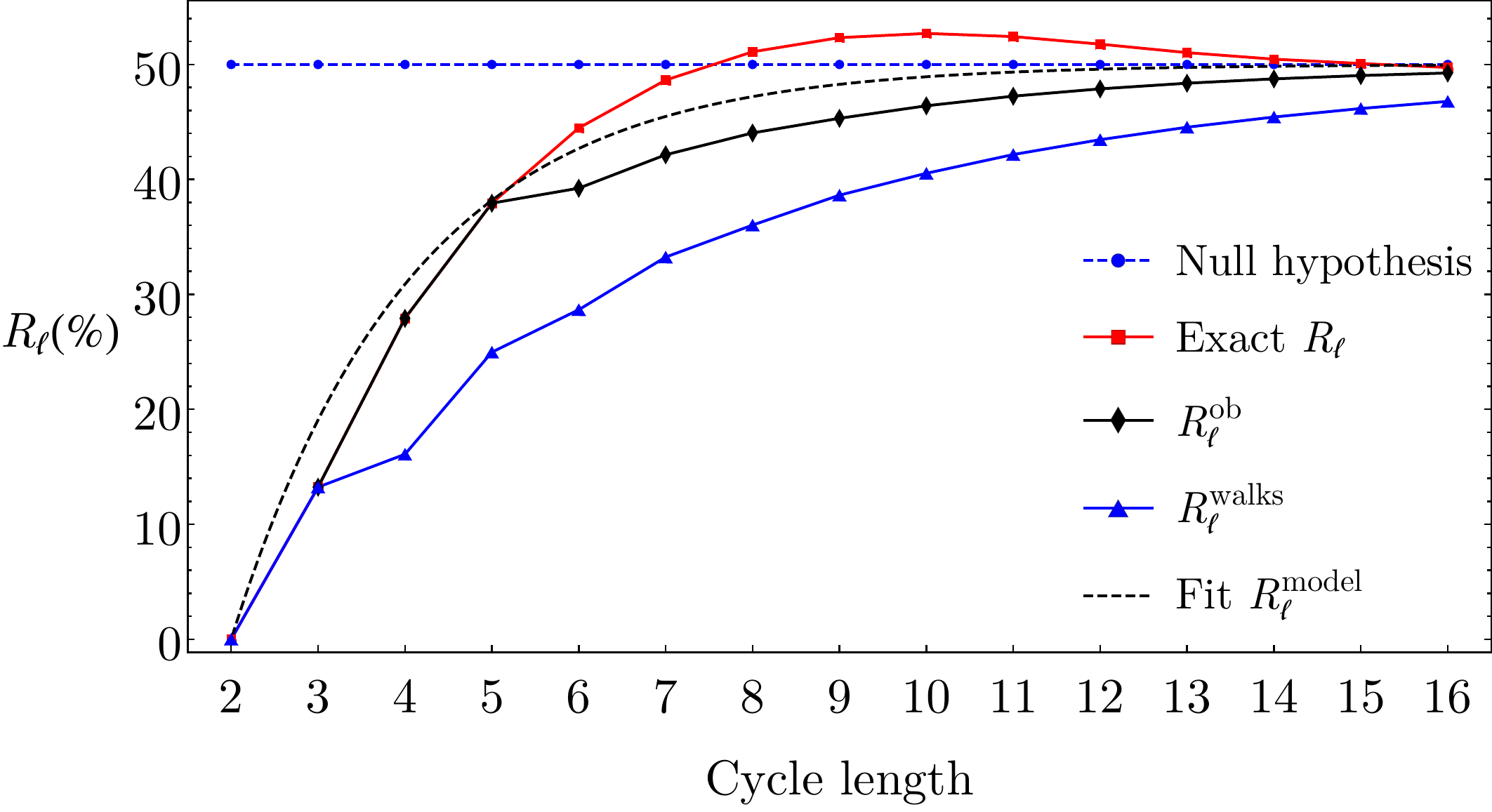}
\end{center}
\vspace{-4mm}
\caption{\label{fig:Gama}Exact percentage $R_\ell$ of negatively signed simple cycles on the Gahuku-Gama network calculated from Eq.~(\ref{CoreEngine}) (red squares) as compared to the null-hypothesis (blue circles). The dashed black curve is a simple exponential fit of $R_\ell$ yielding the correlation length $\xi\simeq 2.08$. Also shown are the percentages $R_{\text{walks};\ell}$ (blue triangles) and $R_{\text{ob};\ell}$ (black diamonds) calculated from the walks and primitive orbits, respectively.}
\end{figure}
The Gahuku-Gama network represents the relation between sixteen tribes living in the eastern central highlands of New-Guinea \cite{Hage1973}. 
Since the network is very small, the Monte Carlo approach is not necessary and we obtained the exact balance from simple cycles of all lengths thanks to Eq.~(\ref{CoreEngine}). The results are shown on Fig.~(\ref{fig:Gama}). Observe how the balance ratios $R^{\text{walks}}_{\ell}$ and $R^{\text{ob}}_{\ell}$ respectively calculated from the walks and primitive orbits  overestimate the proportion of balanced cycles. 

The exact results show that up to length $\ell=7$, the actual ratio $R_\ell$ is well below that of the null-hypothesis, indicating strong inter-edges correlation in favour of balanced cycles. This observation can be made more precise on noting that the balance is well fitted by a simple exponential model  
\begin{equation}\label{Rmodel}
R^{\text{model}}_\ell = \big(1 - e^{-(\ell-2)/2\xi}\big),
\end{equation}
where $\xi \simeq 1.04$ is the correlation length. In this context, $\xi$ characterises the distance between two edges of the graph such that their signs are correlated. Note, the maximum distance between any two vertices on a cycle of length $\ell$ is $\lfloor\ell/2\rfloor$, hence Eq.~(\ref{Rmodel}) fits $2\xi$.
This indicates that tribes of the Gahuku-Gama network are mostly sensitive to the relations with all their first degree neighbours. 
Furthermore, while the network is less balanced than might seem to be the case when considering only the triangles, Fig.~(\ref{fig:Gama}) suggests that much of the imbalance is shifted to long-length simple cycles. In particular, the rebound of $R_\ell$ above $50\%$ for $7<\ell \lesssim 13$ shows that long negative simple cycles are over-represented as compared to a totally uncorrelated model (the null hypothesis). This means that negatively signed situations are slightly favoured in cyclic groups involving $7<\ell \lesssim 13$ tribes. We will  discuss the significance of this observation in the Section~\ref{CompareInterprete}.

~\\

\subsection{WikiElections network}
\begin{figure}[t!]
\begin{center}
\includegraphics[width=.9 \textwidth]{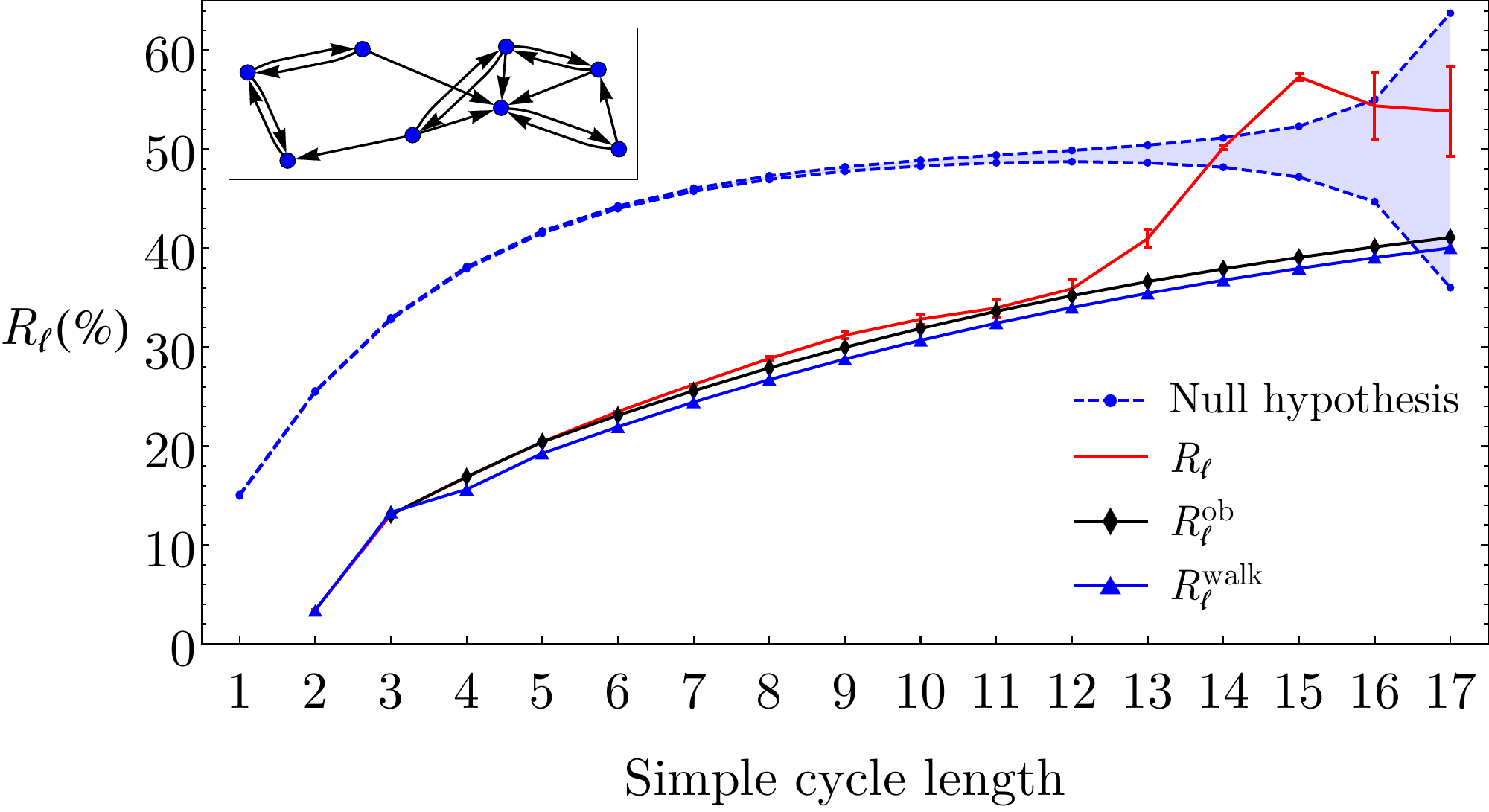}
\end{center}
\vspace{-4mm}
\caption{\label{fig:Wiki}Computed percentage of negatively signed simple cycles on the WikiElection network for cycle length up to 17 (red line and error bars). The blue shaded region bordered by dashed blue lines shows the values of $R_\ell$ compatible with the null-hypothesis, as determined by Eqs.~(\ref{Rnull}, \ref{ErrorNull}). Also shown are the percentages $R^{\text{walks}}_{\ell}$ (blue triangles) and $R^{\text{ob}}_{\ell}$ (black diamonds) calculated from the walks and primitive orbits, respectively. In inset, a subgraph of the WikiElections network.}
\end{figure}
The WikiElections network represents the votes of Wikipedia users during the elections of other users to
adminship. The network is obtained as follows: when a user votes against the candidate, an edge with a negative weight is created from the voting user to the candidate. If instead the user is neutral or supports the candidate, a positive weight is given to this edge. 
The network counts 8,297 vertices and is thus too large for a direct exact calculation of the balance ratios beyond $\ell=6$ and we employed a Monte Carlo approach in tandem with Eq.~(\ref{CoreEngine}). In total we evaluated the balance on 1,800,000 graphs on 20 vertices. The results are shown on Fig.~(\ref{fig:Wiki}).\\[-.5em] 

We find the balance ratio $R_\ell$ to evolve with $\ell$ in three major phases, which we will also observe on the Slashdot and Epinions networks. For short simple cycles $\ell\lesssim12$, $R_\ell$ increases slowly and smoothly with $\ell$ and is also well approximated by the primitive orbits results. In addition, within this range of cycle lengths, $R_\ell$ is always much smaller than predicted by the null-hypothesis, witnessing a strong inter-edge correlation in favour of balance. A sharp transition to $R_\ell$ values circa $50\%$ then occurs around $\ell\sim12-14$. This transition is not an artefact of our algorithm as it is not observed on synthetic networks, see Appendix~\ref{AlgoResultChecks} for details. 

The transition demonstrates that at $\ell\sim12-14$, the simple cycle length becomes longer than twice the inter-edge correlation length $\xi$, which must thus be around $6-7$.\footnote{Indeed, although the network is directed, many of its edges are bidirectional so that the maximum distance between any two vertices on a cycle of length $\ell$ is around $\lfloor\ell/2\rfloor$. We emphasise that $\xi=6-7$ does not mean that individuals participating in the WikiElections network are sensitive to all the relations between their neighbours up the 6th or 7th degree. Rather, $\xi$ only provides an upper bound on the \emph{depth} of the correlation. This is because simple cycles of length $\ell$ typically sustain shortcuts which lower the average distance between the individuals participating in the cycle. The inset of Fig.~\ref{fig:Wiki} illustrates this phenomenon with a subgraph of the WikiElections network sustaining an octagon, but where the average distance between any two vertices is only $\sim 2.5$.} 
Following the sharp transition, $R_\ell$ is reliably found to be over $50\%$, only to slowly decay to results consistent with null-hypothesis \footnote{This effect becomes even more pronounced when $R_\ell$ is compared with the more accurate null model that takes the structural correlations between simple cycles into account.}. This last behaviour, which is also present on the Gahuku-Gama network, suggests that much of the imbalance is shifted to long simple cycles for which edges signs appear to be weakly correlated in favour of imbalance.\\[-.5em]

\subsection{Slashdot network}
\begin{figure}[t!]
\begin{center}
\includegraphics[width=.9 \textwidth]{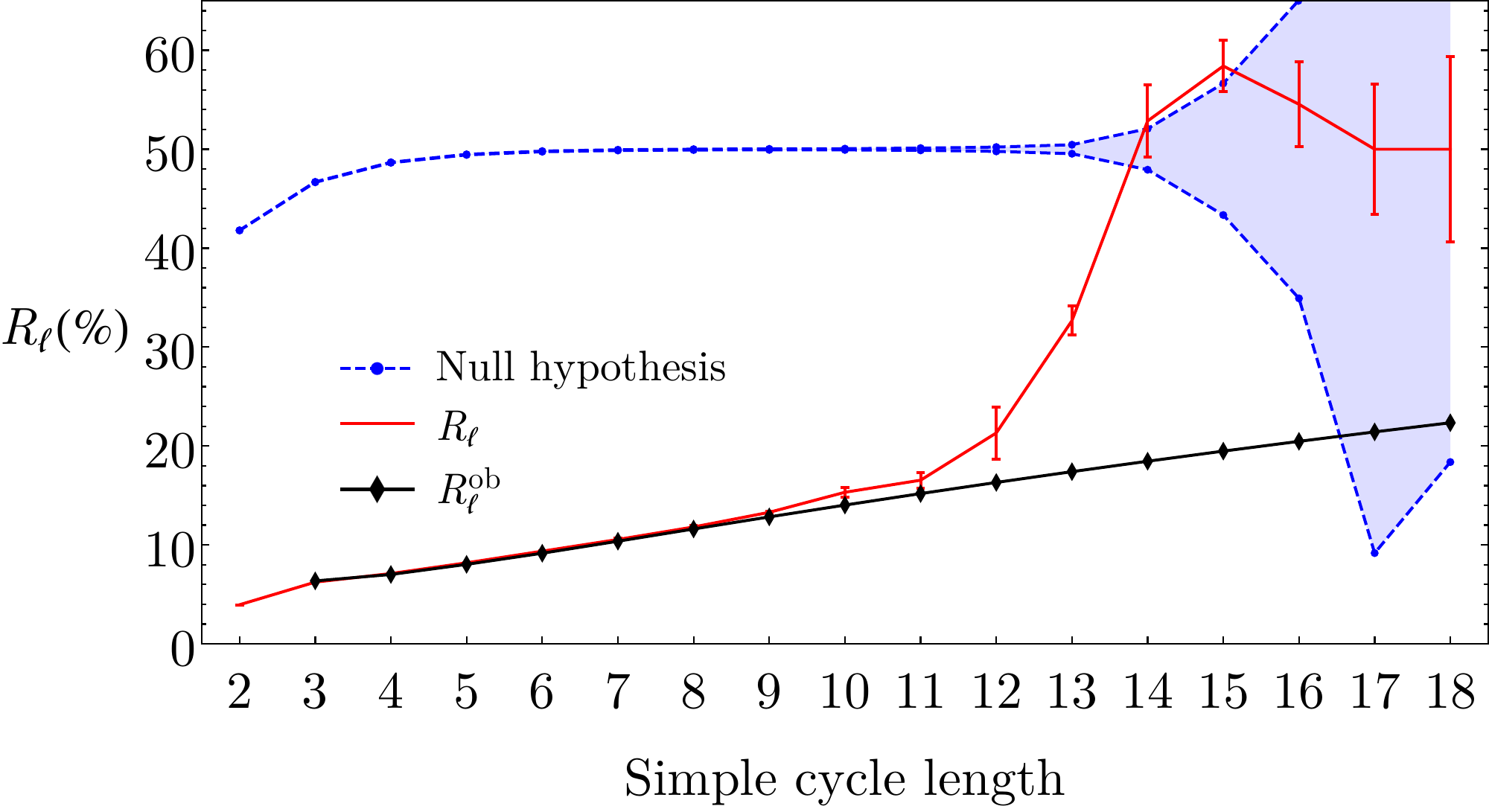}
\end{center}
\vspace{-4mm}
\caption{\label{fig:Slashdot}Computed percentage of negatively signed simple cycles on the Slashdot network for cycle length up to 18 (red line and error bars). The blue shaded region bordered by dashed blue lines shows the values of $R_\ell$ compatible with the null-hypothesis, as determined by Eqs.~(\ref{Rnull}, \ref{ErrorNull}). Also shown is the percentage $R^{\text{ob}}_{\ell}$ determined from the primitive orbits of the graph (black diamonds).}
\end{figure}
The Slashdot network is a large directed graph on 82,144 vertices representing relations of amity/enmity between the users of the Slashdot website \cite{Lampe2007, Kunegis2010}. 
For the Monte Carlo implementation of Eq.~(\ref{CoreEngine}), we sampled 20,000,000 graphs on 20 vertices from this network. We present the balance ratio $R_\ell$ up to $\ell=20$ on Fig.~(\ref{fig:Slashdot}). 

The balance ratio on this network exhibit a behaviour similar to that observed on the WikiElection network: at  first $R_\ell$ increases smoothly with $\ell\leq 10$. Then $R_\ell$ undergoes a rapid transition to higher values broadly consistent with the null hypothesis. This indicates a correlation length between the edges of $\xi\simeq\lfloor11/2\rfloor=5$ and thus a correlation depth of 5 or less.
It is also remarkable that the balance ratios for $14\leq\ell\leq 16$ are once more notably higher than $50\%$, indicating, as in the Gahuku-Gama and WikiElections networks, that much of the imbalance is shifted to long-length cycles. 

\subsection{Epinions}
\begin{figure}[t!]
\begin{center}
\includegraphics[width=.9 \textwidth]{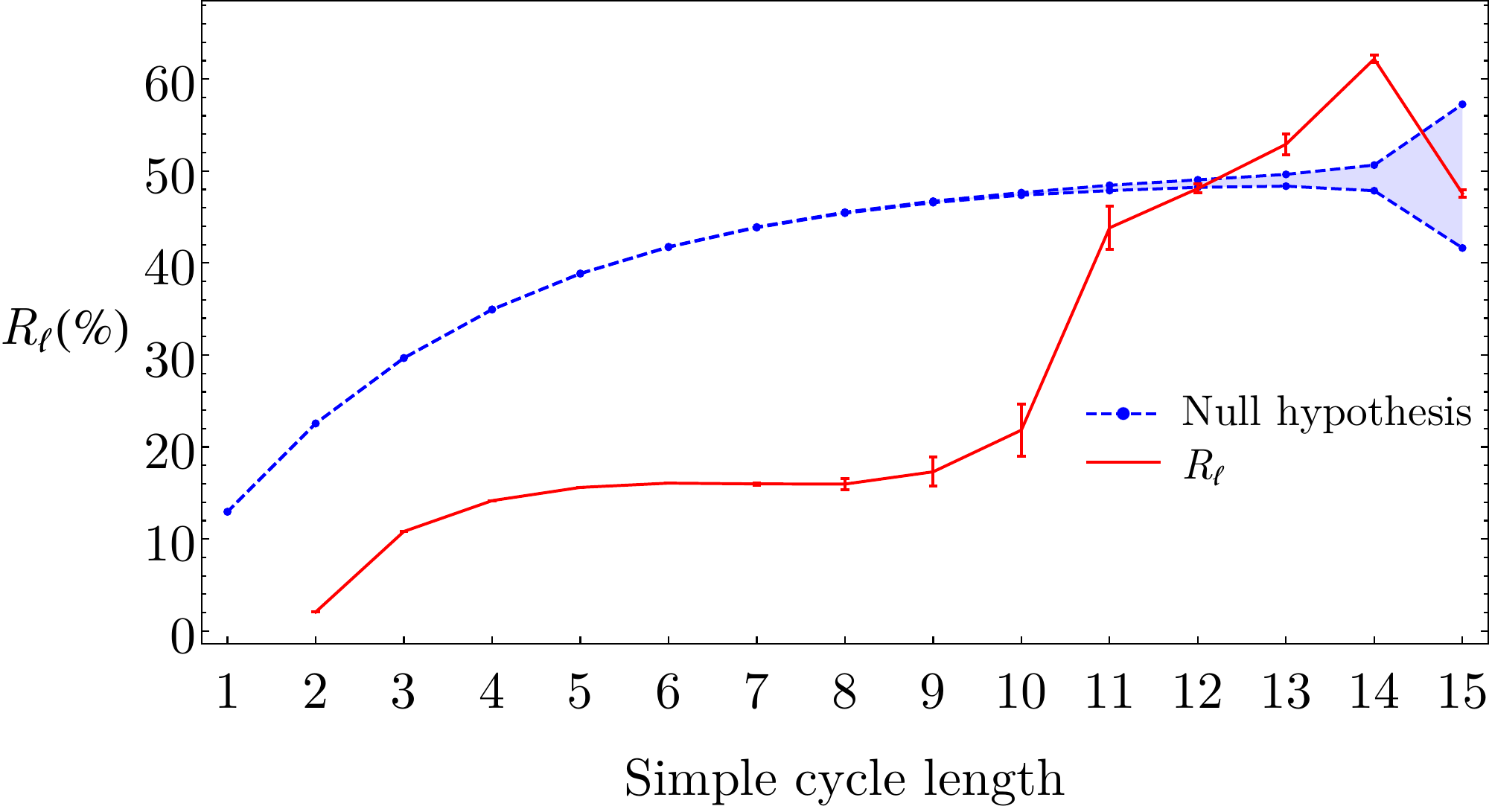}
\end{center}
\vspace{-4mm}
\caption{\label{Fig:Epinions}Computed percentage of negatively signed simple cycles on the Epinions network for cycle length up to 15 (red line and error bars). The blue shaded region bordered by dashed blue lines shows the values of $R_\ell$ compatible with the null-hypothesis, as determined by Eqs.~(\ref{Rnull}, \ref{ErrorNull}).}
\end{figure}

The Epinions network is a large directed graph on 131,828 vertices representing relations between the users of the consumer review website Epinions.com. 
For the Monte Carlo implementation of Eq.~(\ref{CoreEngine}), we sampled 1,000,000,000 (one billion) graphs on 30 vertices from this network. We present the resulting balance ratio $R_\ell$ up to $\ell=15$ on Fig.~(\ref{Fig:Epinions}). 
We were not able to compute the balance ratio $R^{\text{ob}}_{\ell}$ using primitive orbits, owing to the very large size of this network. Indeed, recall that in order to compute the number of primitive orbits, one needs the spectrum of the Hashimoto edge adjacency matrix of the network. This matrix has size $|E|\times |E|$, with $|E|$ the number of (directed) edges of the graph. Counting the primitive orbits thus takes $O(|E|^3)$ time and on Epinions $|E|=841372$. Preliminary calculations indicate that it would take well over a week of computation to get sufficiently many eigenvalues to obtain a reliable approximation of $R^{\text{ob}}_{\ell}$. For this reason, we decided not to seek the primitive orbits results in this case. 

Broadly speaking, the balance ratio $R_\ell$ behaves similarly on this network as it does on the WikiElections and Slashdot ones. The transition of $R_\ell$ from small to high values indicates a correlation length $\xi$ circa $10/2=5$, and thus a correlation depth of 5 or less. Strikingly, for $4\leq \ell\leq 9$, $R_\ell$ is almost constant around $15\%$ witnessing a very strong, almost length-independent, inter-edge correlation.

\subsection{Comparison with past studies and interpretation of the results}\label{CompareInterprete}
Past studies of the WikiElections, Slashdot and Epinions networks have shown them to be very well balanced or quite unbalanced, depending on the measure studied. Kunegis \cite{Kunegis2014} found that these three networks were globally well balanced, with algebraic conflicts on the order of $10^{-3}$ that is a hundred time smaller than in the case of Gama-Gahuku.
Kunegis further confirmed this with the signed clustering and relative signed clustering coefficients, which are sensitive to local structures at the level of triads, and were found to be at least five times smaller on  WikiElections, Slashdot and Epinions than on Gama-Gahuku.
Similar conclusions had been reached by Facchetti \textit{et al.} \cite{Facchetti2011}, who showed using the frustration index that the percentage of perfectly balanced nodes was well over $50\%$ on the three online social networks.\\[-.5em]

At the opposite, Estrada and Benzi \cite{Estrada2014} who studied the degree of balance, argued that these networks are less-well balanced than concluded so far.\footnote{The difference of conclusion might originate from that the degree of balance is sensitive to cycles of all lengths, when the  signed clustering and relative signed clustering coefficients are only influenced by triads. Estrada and Benzi argued that a similar observation could be made for the frustration index, which appeared to follow closely what could be predicted from triads only.}
In particular they found highly unbalanced subnetworks of individuals in WikiElections, Slashdot and Epinions, which we can put in relation with the results presented here. Indeed, Estrada and Benzi predict "the existence of many negative cycles responsible for the global lack of balance" in these networks. They further remark that finding and studying these negative cycles would be extremely expensive computationally. It is tempting to relate the negative cycles predicted by Estrada and Benzi and the cyclic groups of over circa $10$ individuals responsible for the overrepresentation of unbalanced simple cycles at such lengths as visible in our results.

This suggests a possible structural interpretation  for the sharp transition of $R_\ell$ and the ensuing overrepresentation of unbalanced cycles that we have observed. Consider a network comprising well balanced, relatively dense clusters of individuals with negative edges mostly located in between these clusters. Then most short simple cycles will exist inside the dense clusters, consequently appearing strongly balanced. At the opposite, simple cycles whose lengths exceed the typical cluster size, being simple, are forced to cross cluster boundaries, thereby appearing mostly unbalanced. As a corollary, the highly unbalanced subnetworks of individuals are those that include individuals from different well balanced clusters and are visited mostly by long simple cycles. 
This kind of cluster structure can be confirmed in the case of the Gama-Gahuku network. There, Kunegis' in-depth analysis has shown that the network comprises three dense, perfectly balanced clusters with negative edges only between them. The average size of these clusters is 5.3 with the largest one regrouping 7 tribes, in line with our analysis which shows that $R_\ell\gtrsim 50\%$ from $\ell=7$ onwards.  

\section{Conclusion}
\subsection{Balance}
Ever since relations of amity and enmity between individuals have been modelled as positive and negative signs attached to the edges of a network, the possibility of finding patterns in these signs has arisen. 
In one of the earliest formulation of the theory of balance, Heider proposed that most groups of individuals whose relationships form a cycle, should evolve into a state where this cycle has a positive overall sign \cite{Heider1946}.
While much research on the subject has appeared since then, altering the concept of social balance and of Heider's initial conjecture as well as questioning its validity, this very earliest statement remained largely unverified owing to its natural formulation in terms of simple cycles. 
Contrary to closed walks, that is "ordinary" cycles, simple cycles may not visit any vertex more than once and most problems concerning them are difficult, indeed NP-hard and $\#$P-complete.\\[-.5em] 

We thus saw Heider's earliest conjecture as a well posed challenge of network analysis and while its relevance to modern sociology is open for debate, simple cycles undeniably offer several advantages over other cycle-based studies. They are indeed free from the double-counting problem, work on undirected and directed networks alike, and the resulting balance indicators $K_\ell$ and $R_\ell$ take on a range of values, offering a gray-scale measure of the state of the network. Furthermore, distinguishing the contributions of the cycles based on length bridges the divide between local and global balance indices, since e.g. doing so probes edge-correlations on a wide range of length scales. Consequently we sought, as network analysts, to provide the most detailed  possible study of social balance using hitherto unaccessible objects, the simple cycles, via novel combinatorial techniques. We hope that our results will stimulate further research into simple cycles on the one hand, and will be exploited by sociologists in the light of the latest developments in this field on the other.\\[-.5em] 
  
We have studied the proportion $R_\ell$ of negative simple cycles of length $\ell\leq 20$ and its variants $K_\ell$ and $\mathcal{U}_\ell$; so as to be as close as possible to Heider's original conjecture, while also relating our results to the latest research. For exemple, $1-R_3$ equals the triangle index and $K_3$ is the relative signed clustering coefficient. 
Our results show that 
$R_\ell$ is greatly depressed as compared to an independent sampling scenario (the null hypothesis), typically up to lengths of circa $\ell\sim10$. This means that balanced situations are strongly favoured when short simple cycles are considered. 
It is interesting that on the three large networks analysed here (Wikielections, Slashdot and Epinions), a rapid transition from balance (small $R_\ell$ values) to random ($R_\ell\sim R_\ell^{\text{null}}$) occurs around $\ell\sim10$. This is a signature of strong inter-edges correlations with correlation length $\xi\simeq 10/2 = 5$. The correlation depth, which quantifies the degree up to which individuals are correlated with their neighbours, is thus less than or equal to 5. A rebound of the balance ratio over $50\%$ following the transition is also clearly detectable in the data.
This means that unbalanced situations are over-represented for long-length simple cycles as compared to the null-hypothesis.\\[-.5em]

While the relevance of cyclic clusters of individuals in social networks beyond triads is currently controversial, we believe that these do contain sociological information: we observe there is no \emph{mathematical} reason why $R_\ell$ should undergo a sharp transition as seen in the data. In particular, this transition does not exist in the null-hypothesis model, where edge signs are attributed randomly to a real social network, nor does it exist on synthetic networks where both edges and their signs are random. We therefore hope that researchers with a more sociological background than ours will consider the results of this work.  

 
\subsection{Functions on simple cycles and simple paths}
Technically, the approach presented here to study the balance in networks is generally applicable to estimate any function of the simple cycles of a graph. Furthermore, the core combinatorial result of our method immediately extend to vertex-specific questions, e.g. for evaluating the balance of the simple cycles passing through some specified vertex. It also remains valid when asking questions pertaining to simple paths (also known as self-avoiding walks). Both of these observations stem from a matrix extension of Eq.~(\ref{CoreEngine}) which is presented in \cite{Giscard2016b}.
This extension provides a matrix $\mathbf{P}(z)$ whose entry $\mathbf{P}(z)_{ij}$ is the ordinary generating function of the simple paths from $i$ to $j$ ($i\neq j)$ or of the simple-cycles from $i$ to itself ($i=j$). Note, the matrix extension is \emph{not} obtained upon just removing the trace from Eq.~(\ref{CoreEngine}).

This matrix formulation should permit the calculation of such functions as the all-paths kernel \cite{Borgwardt2005} for data-mining algorithms. The all-paths kernel is a positive semi-definite function on pairs of graphs aimed at measuring their degree of similarity via the similarity between their simple paths:
$$
k_{\text{all paths}}(G, G') = \sum_{p~\text{simple path on }G}~~\sum_{p'~\text{simple path on }G'}~k(p,p'),
$$
with $k(p,p')$ a kernel on pairs of paths. Another example is the cyclic pattern kernel \cite{Horvath2004} which evaluates the similarities between the simple cycles of a pair of graphs. Both of these examples currently suffer from efficiency problems which one should overcome (at least partially) with Eq.~(\ref{CoreEngine}) and its matrix extension, possibly in conjunction with a Monte Carlo approach as effected here.

\section*{Acknowledgment}
This work was supported by Royal Commission for the Exhibition of 1851 [research fellowship to P.-L. G.] We thank two anonymous referees for their numerous and detailed comments which have helped improve this paper.

\appendix
\section{Checking the method against bias}\label{AlgoResultChecks}
While the convergence of our calculations can be assessed via the standard deviation of the results, Monte Carlo approaches are not immune to bias. It is thus necessary to verify the quality of the results independently of the method itself: 
\begin{itemize}
\item[i)] We computed the exact balance ratios $R_1$, $R_2$ and $R_3$ via conventional means and verified our results to be consistent with these, see Appendix~\ref{FullResults}, Tables~\ref{Wikiresults}, \ref{Slashdotresults} and \ref{EpinionsTable}. In the case of WikiElections, we also computed $R_4$, $R_5$ and $R_6$ exactly using Eq.~(\ref{CoreEngine}).
\item[ii)] When available, we used the primitive orbits results to verify that the balance ratios $R_4$ and $R_5$ predicted by the algorithm were consistent with the exact results. Indeed, recall that up to $\ell=5$, $R^{\text{ob}}_\ell = R_\ell$ exactly. 
\item[iii)] On the Gahuku-Gama network, we verified that the Monte Carlo results are consistent with the exact balance ratios at all lengths. 
\end{itemize}

In addition, we found the WikiElections, Slashdot and Epinions networks to exhibit sharp transitions of their balance ratios from low values $R_\ell\sim 10-30\%$ up to values consistent with the null-hypothesis circa $50\%$. Given the importance of this observation, it is necessary to check that it is not an artifact of the algorithm we employed: 
\begin{itemize}
\item[i)] In the case of the WikiElections network, we reallocated the edge signs randomly and ran our approach on the resulting signed graph. The balance ratios did not exhibit any sharp transition anymore but rather were consistent with the more accurate null model. This indicates that the transition is not an artifact of our method. 
\item[ii)] We should expect a transition of $R_\ell$ to values consistent with the null hypothesis as the cycle length becomes longer than the correlation length. Indeed, if there was no sharp transition, a simple extrapolation of the trend exhibited by the first five (exactly known) balance ratios, suggests that $R_\ell$ would not be reach 50\% until at least $\ell\gtrsim 50$. This conservative estimate would mean that $\xi\gtrsim 25$ or more, a number that is far too large to be plausible.
\end{itemize}

\section{Proof of Proposition \ref{PropNT}}\label{Proof}
The proposition results from equating the product and determinant forms of the Ihara zeta function. Recall that
\begin{equation*}
\zeta^{-1}_{|G|}(z)=\prod_{c\in[C]} \left(1-z^{\ell(c)}\right)= \prod_{\ell} \left(1-z^{\ell}\right)^{N_{\text{ob};\,\ell}},
\end{equation*}
where $N_{\text{ob};\,\ell}=N_{\text{ob};\,\ell}^++N_{\text{ob};\,\ell}^-$ is the total number of primitive orbits of length $\ell$. Thus we have
\begin{equation}
\zeta^{-1}_{|G|}(z)=\prod_{\ell} \left(1-z^{\ell}\right)^{N_{\text{ob};\,\ell}}=\det\left(\mathbf{I}-z|\mathbf{T}|\right).
\end{equation}
Now, on taking the logarithm on both sides we obtain
$$
\sum_{i=1}^\infty\frac{1}{i}\,z^i\, \operatorname{Tr}|\mathbf{T}|^i=\sum_{j}^\infty\sum_{k=1}^\infty N_{\text{ob};\,j}\frac{z^{kj}}{k}.
$$
Equating the coefficient of $z^\ell$ on left and right hand sides then gives
$$
\frac{1}{\ell}\operatorname{Tr}|\mathbf{T}|^\ell=\sum_{k|\ell}^\infty\frac{1}{k}N_{\text{ob};\,\ell/k},
$$
and a M\"{o}bius inversion finally provides $N_{\text{ob};\ell}$
$$
N_{\text{ob};\,\ell}^++N_{\text{ob};\,\ell}^- = \frac{1}{\ell}\sum_{k|\ell}\mu(\ell/k) \operatorname{Tr}
|\mathbf{T}|^k.
$$
The proof is entirely similar on signed networks, where $\mathbf{T}$ replaces $|\mathbf{T}|$ and  $N_{\text{ob};\,\ell}^+-N_{\text{ob};\,\ell}^-$ is obtained instead of $N_{\text{ob};\,\ell}^++N_{\text{ob};\,\ell}^-$.\qed

%

\section{Full numerical results}\label{FullResults}
In this section we present the full numerical results obtained on the four networks mentioned earlier, which include the balance ratios, $R_\ell = (N_\ell^-)/(N_\ell^-+N_\ell^+)$, ratios of negative to positive simple cycles
$
\mathcal{U}_\ell = N_\ell^-/N_\ell^+
$
and the relative signed clustering coefficient $K_\ell := (N_\ell^+-N_\ell^-)/(N_\ell^-+N_\ell^+)$  up to $\ell=20$. 
We  also give the exact values of $R_\ell$, $\mathcal{U}_\ell$ and $K_\ell$ for the self-loops, backtracks and triangles ($\ell=1,2,3$). These are respectively given by
\begin{align*}
R_1 &= \frac{\operatorname{Tr}(|\mathbf{A}|-\mathbf{A})}{2\operatorname{Tr}|\mathbf{A}|},\quad R_2 = \frac{\operatorname{Tr}(|\tilde{\mathbf{A}}|^2-\tilde{\mathbf{A}}^2)}{2\operatorname{Tr}|\tilde{\mathbf{A}}|^2},\quad R_3 = \frac{\operatorname{Tr}(|\tilde{\mathbf{A}}|^3-\tilde{\mathbf{A}}^3)}{2\operatorname{Tr}|
\tilde{\mathbf{A}}|^3},\\
\mathcal{U}_1 &= \frac{\operatorname{Tr}(|\mathbf{A}|-\mathbf{A})}{\operatorname{Tr}(|\mathbf{A}|+\mathbf{A})},\hspace{2.7mm} \mathcal{U}_2 = \frac{\operatorname{Tr}(|\tilde{\mathbf{A}}|^2-\tilde{\mathbf{A}}^2)}{\operatorname{Tr}(|\tilde{\mathbf{A}}|^2+\tilde{\mathbf{A}}^2)},\hspace{3mm} \mathcal{U}_3 = \frac{\operatorname{Tr}(|\tilde{\mathbf{A}}|^3-\tilde{\mathbf{A}}^3)}{\operatorname{Tr}(|\tilde{\mathbf{A}}|^3+\tilde{\mathbf{A}}^3)},\\
K_1 &= \frac{\operatorname{Tr}(\mathbf{A})}{\operatorname{Tr}(|\mathbf{A}|)},\hspace{9.7mm} K_2 = \frac{\operatorname{Tr}(\tilde{\mathbf{A}}^2)}{\operatorname{Tr}(|\tilde{\mathbf{A}}|^2)},\hspace{11.1mm}  K_3 = \frac{\operatorname{Tr}(\tilde{\mathbf{A}}^3)}{\operatorname{Tr}(|\tilde{\mathbf{A}}|^3)},
\end{align*}
where $\tilde{\mathbf{A}} = \mathbf{A}-\text{Diag}(\mathbf{A})$.

\begin{table*}[h!]
\renewcommand{\arraystretch}{1.1}
\resizebox{\textwidth}{!}{
\begin{tabular}{|c|c|c|c|c|c|}
\hline
\hline  
&\textbf{Self-loops $R_1$}& \textbf{Backtracks $R_2$} & \textbf{Triangles $R_3$}& \textbf{Squares $R_4$}& \textbf{Pentagons $R_5$}\\
\hline
\textbf{Exact} & &$0\%$&$13.24\%$&$27.92\%$&$37.93\%$\\
 \hline 
 \multicolumn{6}{c}{}\\
 \hline
& \textbf{ $R_6$}&  \textbf{ $R_7$}& \textbf{ $R_8$}& \textbf{ $R_9$}& \textbf{ $R_{10}$}\\
\hline
\textbf{Exact} &$44.47\%$&$48.64\%$&$51.10\%$&$52.33\%$&$52.7\%$\\ 
 \hline 
 \multicolumn{6}{c}{}\\
 \hline
& \textbf{ $R_{11}$}&  \textbf{ $R_{12}$}& \textbf{ $R_{13}$}& \textbf{ $R_{14}$}& \textbf{ $R_{15}$}\\
\hline
\textbf{Exact} &$52.43\%$&$51.77\%$&$51.03\%$&$50.46\%$&$50.09\%$\\ 
 \hline 
 \multicolumn{6}{c}{}\\
 \hline
& \textbf{ $R_{16}$}&  \textbf{ $R_{17}$}& \textbf{ $R_{18}$}& \textbf{ $R_{19}$}& \textbf{ $R_{20}$}\\
\hline
\textbf{Exact} &$49.74\%$&&&& \\ 
\hline
\hline
\end{tabular}}
\vspace{-2mm}
\caption{\label{Gamaresults}Exact balance ratios $R_\ell$ for $1\leq\ell\leq 20$ obtained on the Gama network using Eq.~(\ref{CoreEngine}).}
\end{table*}

\begin{table*}[h!]
\renewcommand{\arraystretch}{1.1}
\resizebox{\textwidth}{!}{
\begin{tabular}{|c|c|c|c|c|c|}
\hline
\hline  
&\textbf{Self-loops $\mathcal{U}_1$}& \textbf{Backtracks $\mathcal{U}_2$} & \textbf{Triangles $\mathcal{U}_3$}& \textbf{Squares $\mathcal{U}_4$}& \textbf{Pentagons $\mathcal{U}_5$}\\
\hline
\textbf{Exact} & &$0\%$&$15.3\%$&$38.7\%$&$61.1\%$\\
 \hline 
 \multicolumn{6}{c}{}\\
 \hline
& \textbf{ $\mathcal{U}_6$}&  \textbf{ $\mathcal{U}_7$}& \textbf{ $\mathcal{U}_8$}& \textbf{ $\mathcal{U}_9$}& \textbf{ $\mathcal{U}_{10}$}\\
\hline
\textbf{Exact} &$80.1\%$&$94.7\%$&$104.5\%$&$109.8\%$&$111.4\%$\\ 
 \hline 
 \multicolumn{6}{c}{}\\
 \hline
& \textbf{ $\mathcal{U}_{11}$}&  \textbf{ $\mathcal{U}_{12}$}& \textbf{ $\mathcal{U}_{13}$}& \textbf{ $\mathcal{U}_{14}$}& \textbf{ $\mathcal{U}_{15}$}\\
\hline
\textbf{Exact} &$110.2\%$&$107.3\%$&$104.2\%$&$101.9\%$&$100.3\%$\\ 
 \hline 
 \multicolumn{6}{c}{}\\
 \hline
& \textbf{ $\mathcal{U}_{16}$}&  \textbf{ $\mathcal{U}_{17}$}& \textbf{ $\mathcal{U}_{18}$}& \textbf{ $\mathcal{U}_{19}$}& \textbf{ $\mathcal{U}_{20}$}\\
\hline
\textbf{Exact} &$99\%$&&&& \\ 
\hline
\hline
\end{tabular}}
\vspace{-2mm}
\caption{\label{Gamaresults2}Exact ratio of negative to positive simple cycles $\mathcal{U}_\ell$ for $1\leq\ell\leq 20$ obtained on the Gama network using Eq.~(\ref{CoreEngine}).}
\end{table*}

\begin{table*}[h!]
\renewcommand{\arraystretch}{1.1}
\resizebox{\textwidth}{!}{
\begin{tabular}{|c|c|c|c|c|c|}
\hline
\hline  
&\textbf{Self-loops $K_1$}& \textbf{Backtracks $K_2$} & \textbf{Triangles $K_3$}& \textbf{Squares $K_4$}& \textbf{Pentagons $K_5$}\\
\hline
\textbf{Exact} & &$1$&$0.735$&$0.442$&$0.241$\\
 \hline 
 \multicolumn{6}{c}{}\\
 \hline
& \textbf{ $K_6$}&  \textbf{ $K_7$}& \textbf{ $K_8$}& \textbf{ $K_9$}& \textbf{ $K_{10}$}\\
\hline
\textbf{Exact} &$0.111$&$0.027$&$-0.022$&$-0.047$&$-0.054$\\ 
 \hline 
 \multicolumn{6}{c}{}\\
 \hline
& \textbf{ $K_{11}$}&  \textbf{ $K_{12}$}& \textbf{ $K_{13}$}& \textbf{ $K_{14}$}& \textbf{ $K_{15}$}\\
\hline
\textbf{Exact} &$-0.049$&$-0.035$&$-0.021$&$-0.009$&$-0.002\%$\\ 
 \hline 
 \multicolumn{6}{c}{}\\
 \hline
& \textbf{ $K_{16}$}&  \textbf{ $K_{17}$}& \textbf{ $K_{18}$}& \textbf{ $K_{19}$}& \textbf{ $K_{20}$}\\
\hline
\textbf{Exact} &$0.005$&&&& \\ 
\hline
\hline
\end{tabular}}
\vspace{-2mm}
\caption{\label{Gamaresults3}Exact relative signed cluster coefficient $K_\ell$ for $1\leq\ell\leq 20$ obtained on the Gama network using Eq.~(\ref{CoreEngine}). Recall that $-1\leq K_\ell\leq 1$ and that $K_\ell=1$ designates a totally balanced situation, while $K_\ell=-1$ characterises a totally unbalanced one.}
\end{table*}

\begin{table*}[h!]
\renewcommand{\arraystretch}{1.1}
\resizebox{\textwidth}{!}{
\begin{tabular}{|c|c|c|c|c|c|}
\hline
\hline  
&\textbf{Self-loops $R_1$}& \textbf{Backtracks $R_2$} & \textbf{Triangles $R_3$}& \textbf{Squares $R_4$}& \textbf{Pentagons $R_5$}\\
\hline
\textbf{Monte Carlo}  &$45.488\pm0.048\%$ &$3.436\pm0.006\%$&$13.075\pm0.014\%$&$16.862\pm0.098\%$&$20.421\pm0.052\%$\\
\hline 
 \textbf{Exact}                  &$45.455\%$  & $3.438\%$ & $13.068\%$&$16.875\%$&$20.393\%$\\
 \hline 
 \multicolumn{6}{c}{}\\
 \hline
& \textbf{ $R_6$}&  \textbf{ $R_7$}& \textbf{ $R_8$}& \textbf{ $R_9$}& \textbf{ $R_{10}$}\\
\hline
\textbf{Monte Carlo}&$23.43\pm0.05\%$&$26.22\pm0.03\%$&$28.84\pm0.19\%$&$31.2\pm0.31\%$&$32.82\pm0.54\%$\\ 
\hline 
 \textbf{Exact}                  &$23.401\%$  & & &&\\
 \hline 
 \multicolumn{6}{c}{}\\
 \hline
& \textbf{ $R_{11}$}&  \textbf{ $R_{12}$}& \textbf{ $R_{13}$}& \textbf{ $R_{14}$}& \textbf{ $R_{15}$}\\
\hline
\textbf{Monte Carlo}&$33.95\pm0.9\%$&$35.89\pm0.9\%$&$40.96\pm0.9\%$&$50.16\pm0.19\%$&$57.3\pm0.36\%$ \\ 
 \hline 
 \multicolumn{6}{c}{}\\
 \hline
& \textbf{ $R_{16}$}&  \textbf{ $R_{17}$}& \textbf{ $R_{18}$}& \textbf{ $R_{19}$}& \textbf{ $R_{20}$}\\
\hline
\textbf{Monte Carlo} &$54.38\pm3.41\%$&$53.85\pm4.57\%$&&& \\ 
\hline
\hline
\end{tabular}}
\vspace{-2mm}
\caption{\label{Wikiresults}Computed balance ratios $R_\ell$ for $1\leq\ell\leq 20$ on the WikiElections network together with twice the standard deviation exhibited by the Monte Carlo results. We found no simple cycle of length 18, 19 or 20 on this network.
}
\end{table*}

\begin{table*}[h!]
\renewcommand{\arraystretch}{1.1}
\resizebox{\textwidth}{!}{
\begin{tabular}{|c|c|c|c|c|c|}
\hline
\hline  
&\textbf{Self-loops $\mathcal{U}_1$}& \textbf{Backtracks $\mathcal{U}_2$} & \textbf{Triangles $\mathcal{U}_3$}& \textbf{Squares $\mathcal{U}_4$}& \textbf{Pentagons $\mathcal{U}_5$}\\
\hline
\textbf{Monte Carlo}  &$83.45\pm0.17\%$ &$3.56\pm0.006\%$&$15.04\pm0.02\%$&$20.3\pm0.14\%$&$25.7\pm0.08\%$\\
\hline 
 \textbf{Exact}                  &$83.33\%$  & $3.56\%$ & $15.03\%$&&\\
 \hline 
 \multicolumn{6}{c}{}\\
 \hline
& \textbf{ $\mathcal{U}_6$}&  \textbf{ $\mathcal{U}_7$}& \textbf{ $\mathcal{U}_8$}& \textbf{ $\mathcal{U}_9$}& \textbf{ $\mathcal{U}_{10}$}\\
\hline
\textbf{Monte Carlo}&$30.7\pm0.05\%$&$35.5\pm0.06\%$&$40.5\pm0.37\%$&$45.3\pm0.7\%$&$48.9\pm1.2\%$\\ 
 \hline 
 \multicolumn{6}{c}{}\\
 \hline
& \textbf{ $\mathcal{U}_{11}$}&  \textbf{ $\mathcal{U}_{12}$}& \textbf{ $\mathcal{U}_{13}$}& \textbf{ $\mathcal{U}_{14}$}& \textbf{ $\mathcal{U}_{15}$}\\
\hline
\textbf{Monte Carlo}&$51.4\pm2\%$&$56\pm2.2\%$&$69.4\pm2.5\%$&$100.7\pm0.8\%$&$134.2\pm2\%$ \\ 
 \hline 
 \multicolumn{6}{c}{}\\
 \hline
& \textbf{ $\mathcal{U}_{16}$}&  \textbf{ $\mathcal{U}_{17}$}& \textbf{ $\mathcal{U}_{18}$}& \textbf{ $\mathcal{U}_{19}$}& \textbf{ $\mathcal{U}_{20}$}\\
\hline
\textbf{Monte Carlo} &$119\pm15\%$&$116\pm20\%$&&& \\ 
\hline
\hline
\end{tabular}}
\vspace{-2mm}
\caption{\label{Wikiresults2}Computed ratio of negative to positive simple cycles $\mathcal{U}_\ell$ for $1\leq\ell\leq 20$ on the WikiElections network together with twice the standard deviation exhibited by the Monte Carlo results. We found no simple cycle of length 18, 19 or 20 on this network.
}
\end{table*}

\begin{table*}[h!]
\renewcommand{\arraystretch}{1.1}
\resizebox{\textwidth}{!}{
\begin{tabular}{|c|c|c|c|c|c|}
\hline
\hline  
&\textbf{Self-loops $K_1$}& \textbf{Backtracks $K_2$} & \textbf{Triangles $K_3$}& \textbf{Squares $K_4$}& \textbf{Pentagons $K_5$}\\
\hline
\textbf{Monte Carlo}  &$0.0902\pm0.0014$ &$0.9312\pm0.0001$&$0.7385\pm0.0004$&$0.663\pm0.003$&$0.592\pm0.001$\\
\hline 
 \textbf{Exact}                  &$0.0909$  & $0.9312$ & $0.7386$&&\\
 \hline 
 \multicolumn{6}{c}{}\\
 \hline
& \textbf{ $K_6$}&  \textbf{ $K_7$}& \textbf{ $K_8$}& \textbf{ $K_9$}& \textbf{ $K_{10}$}\\
\hline
\textbf{Monte Carlo}&$0.53\pm0.001$&$0.476\pm0.001$&$0.423\pm0.005$&$0.376\pm0.009$&$0.344\pm0.015$\\ 
 \hline 
 \multicolumn{6}{c}{}\\
 \hline
& \textbf{ $K_{11}$}&  \textbf{ $K_{12}$}& \textbf{ $K_{13}$}& \textbf{ $K_{14}$}& \textbf{ $K_{15}$}\\
\hline
\textbf{Monte Carlo}&$0.321\pm0.025$&$0.282\pm0.025$&$0.181\pm0.025$&$-0.003\pm0.005$&$-0.146\pm0.01$ \\ 
 \hline 
 \multicolumn{6}{c}{}\\
 \hline
& \textbf{ $K_{16}$}&  \textbf{ $K_{17}$}& \textbf{ $K_{18}$}& \textbf{ $K_{19}$}& \textbf{ $K_{20}$}\\
\hline
\textbf{Monte Carlo} &$-0.088\pm0.096$&$-0.077\pm0.129$&&& \\ 
\hline
\hline
\end{tabular}}
\vspace{-2mm}
\caption{\label{Wikiresults3}Computed relative signed cluster coefficient $K_\ell$ for $1\leq\ell\leq 20$ on the WikiElections network together with twice the standard deviation exhibited by the Monte Carlo results.
}
\end{table*}

\begin{table*}[h!]
\renewcommand{\arraystretch}{1.1}
\resizebox{\textwidth}{!}{
\begin{tabular}{|c|c|c|c|c|c|}
\hline
\hline  
&\textbf{Self-loops $R_1$}& \textbf{Backtracks $R_2$} & \textbf{Triangles $R_3$}& \textbf{Squares $R_4$}& \textbf{Pentagons $R_5$}\\
\hline
\textbf{Monte Carlo}  & &$3.9995\pm0.001\%$&$6.352\pm0.017\%$&$7.108\pm0.013\%$&$8.192\pm0.013\%$\\
\hline 
 \textbf{Exact}                  & & 4.0003$\%$                & 6.3608$\%$ &&\\
 \hline 
 \multicolumn{6}{c}{}\\
 \hline
& \textbf{ $R_6$}&  \textbf{ $R_7$}& \textbf{ $R_8$}& \textbf{ $R_9$}& \textbf{ $R_{10}$}\\
\hline
\textbf{Monte Carlo} &$9.37\pm0.06\%$&$10.55\pm0.12\%$&$11.82\pm0.15\%$&$13.3\pm0.09\%$&$15.32\pm0.46\%$\\ 
 \hline 
 \multicolumn{6}{c}{}\\
 \hline
& \textbf{ $R_{11}$}&  \textbf{ $R_{12}$}& \textbf{ $R_{13}$}& \textbf{ $R_{14}$}& \textbf{ $R_{15}$}\\
\hline
\textbf{Monte Carlo} &$16.5\pm0.8\%$&$21.3\pm2.6\%$&$32.7\pm1.5\%$&$52.9\pm3.6\%$&$58.4\pm2.6\%$\\ 
 \hline 
 \multicolumn{6}{c}{}\\
 \hline
& \textbf{ $R_{16}$}&  \textbf{ $R_{17}$}& \textbf{ $R_{18}$}& \textbf{ $R_{19}$}& \textbf{ $R_{20}$}\\
\hline
\textbf{Monte Carlo} & $54.5\pm4.3\%$&$50.0\pm6.6\%$&$50.0\pm9.4\%$&&\\ 
\hline
\hline
\end{tabular}}
\vspace{-2mm}
\caption{\label{Slashdotresults}Computed balance ratios $R_\ell$ for $1\leq\ell\leq 20$ on the Slashdot network together with twice the standard deviation exhibited by the Monte Carlo results. We found no simple cycle of length 19 or 20 on this network. 
}
\end{table*}
\begin{table*}[h!]
\renewcommand{\arraystretch}{1.1}
\resizebox{\textwidth}{!}{
\begin{tabular}{|c|c|c|c|c|c|}
\hline
\hline  
&\textbf{Self-loops $\mathcal{U}_1$}& \textbf{Backtracks $\mathcal{U}_2$} & \textbf{Triangles $\mathcal{U}_3$}& \textbf{Squares $\mathcal{U}_4$}& \textbf{Pentagons $R_5$}\\
\hline
\textbf{Monte Carlo}  & &$4.166\pm0.001\%$&$6.783\pm0.019\%$&$7.552\pm0.015\%$&$8.923\pm0.015\%$\\
\hline 
 \textbf{Exact}                  & & 4.167$\%$                & 6.793$\%$ &&\\
 \hline 
 \multicolumn{6}{c}{}\\
 \hline
& \textbf{ $\mathcal{U}_6$}&  \textbf{ $\mathcal{U}_7$}& \textbf{ $\mathcal{U}_8$}& \textbf{ $\mathcal{U}_9$}& \textbf{ $\mathcal{U}_{10}$}\\
\hline
\textbf{Monte Carlo} &$10.34\pm0.07\%$&$11.79\pm0.15\%$&$13.40\pm0.19\%$&$15.3\pm0.12\%$&$18.09\pm0.64\%$\\ 
 \hline 
 \multicolumn{6}{c}{}\\
 \hline
& \textbf{ $\mathcal{U}_{11}$}&  \textbf{ $\mathcal{U}_{12}$}& \textbf{ $\mathcal{U}_{13}$}& \textbf{ $\mathcal{U}_{14}$}& \textbf{ $\mathcal{U}_{15}$}\\
\hline
\textbf{Monte Carlo} &$19.8\pm1.1\%$&$27.1\pm4.1\%$&$48.6\pm3.2\%$&$112.3\pm6.7\%$&$140\pm15\%$\\ 
 \hline 
 \multicolumn{6}{c}{}\\
 \hline
& \textbf{ $\mathcal{U}_{16}$}&  \textbf{ $\mathcal{U}_{17}$}& \textbf{ $\mathcal{U}_{18}$}& \textbf{ $\mathcal{U}_{19}$}& \textbf{ $\mathcal{U}_{20}$}\\
\hline
\textbf{Monte Carlo} & $120\pm20\%$&$100.0\pm25\%$&$100\pm31\%$&&\\ 
\hline
\hline
\end{tabular}}
\vspace{-2mm}
\caption{\label{Slashdotresults2}Computed ratio of negative to positive simple cycles $\mathcal{U}_\ell$ for $1\leq\ell\leq 20$ on the Slashdot network together with twice the standard deviation exhibited by the Monte Carlo results. We found no simple cycle of length 19 or 20 on this network.}
\end{table*}
\begin{table*}[h!]
\renewcommand{\arraystretch}{1.1}
\resizebox{\textwidth}{!}{
\begin{tabular}{|c|c|c|c|c|c|}
\hline
\hline  
&\textbf{Self-loops $K_1$}& \textbf{Backtracks $K_2$} & \textbf{Triangles $K_3$}& \textbf{Squares $K_4$}& \textbf{Pentagons $R_5$}\\
\hline
\textbf{Monte Carlo}  & &$0.92001\pm0.00003$&$0.8730\pm0.0005$&$0.8578\pm0.0004$&$0.8362\pm0.0004$\\
\hline 
 \textbf{Exact}                  & & $0.92000$                & $0.8728$ &&\\
 \hline 
 \multicolumn{6}{c}{}\\
 \hline
& \textbf{ $K_6$}&  \textbf{ $K_7$}& \textbf{ $K_8$}& \textbf{ $K_9$}& \textbf{ $K_{10}$}\\
\hline
\textbf{Monte Carlo} &$0.813\pm0.002$&$0.789\pm0.003$&$0.764\pm0.004$&$0.734\pm0.003$&$0.694\pm0.013$\\ 
 \hline 
 \multicolumn{6}{c}{}\\
 \hline
& \textbf{ $K_{11}$}&  \textbf{ $K_{12}$}& \textbf{ $K_{13}$}& \textbf{ $K_{14}$}& \textbf{ $K_{15}$}\\
\hline
\textbf{Monte Carlo} &$0.67\pm0.023$& $0.574\pm0.074$& $0.346\pm0.042$&     $-0.058\pm0.102$&$-0.168\pm0.074$\\ 
 \hline 
 \multicolumn{6}{c}{}\\
 \hline
& \textbf{ $K_{16}$}&  \textbf{ $K_{17}$}& \textbf{ $K_{18}$}& \textbf{ $K_{19}$}& \textbf{ $K_{20}$}\\
\hline
\textbf{Monte Carlo} & $-0.09\pm0.122$&$0\pm0.187$&$0.001\pm0.266$&&\\ 
\hline
\hline
\end{tabular}}
\vspace{-2mm}
\caption{\label{Slashdotresults3}Computed relative signed cluster coefficient $K_\ell$ for $1\leq\ell\leq 20$ on the Slashdot network together with twice the standard deviation exhibited by the Monte Carlo results.}
\end{table*}

\begin{table*}[h!]
\renewcommand{\arraystretch}{1.1}
\resizebox{\textwidth}{!}{
\begin{tabular}{|c|c|c|c|c|c|}
\hline
\hline  
&\textbf{Self-loops $R_1$}& \textbf{Backtracks $R_2$} & \textbf{Triangles $R_3$}& \textbf{Squares $R_4$}& \textbf{Pentagons $R_5$}\\
\hline
\textbf{Monte Carlo}  & $6.1082\pm0.0002\%$& $2.0857\pm0.0003\%$ & $11.2355\pm0.0016\%$ &$ 14.17\pm0.002\%$ &$15.61\pm0.007\%$\\
\hline 
 \textbf{Exact}                  & $6.1082\%$ & $2.0858\%$ & $11.2343\%$ &  & \\
 \hline 
 \multicolumn{6}{c}{}\\
 \hline
& \textbf{ $R_6$}&  \textbf{ $R_7$}& \textbf{ $R_8$}& \textbf{ $R_9$}& \textbf{ $R_{10}$}\\
\hline
\textbf{Monte Carlo} & $16.07\pm0.02\%$ &$15.99\pm0.12\%$ &$15.97\pm0.58\%$& $17.30\pm1.58\%$&$21.83\pm2.84\%$\\ 
 \hline 
 \multicolumn{6}{c}{}\\
 \hline
& \textbf{ $R_{11}$}&  \textbf{ $R_{12}$}& \textbf{ $R_{13}$}& \textbf{ $R_{14}$}& \textbf{ $R_{15}$}\\
\hline
\textbf{Monte Carlo} & $43.8 \pm 2.3\%$ &$48.1\pm0.5\%$ &$52.9\pm1.2\%$&$62.2\pm0.4\%$&$47.6\pm0.4\%$\\ 
 \hline 
 \multicolumn{6}{c}{}\\
 \hline
& \textbf{ $R_{16}$}&  \textbf{ $R_{17}$}& \textbf{ $R_{18}$}& \textbf{ $R_{19}$}& \textbf{ $R_{20}$}\\
\hline
\textbf{Monte Carlo} &  &$50\%$ && $44.4\%$&\\ 
\hline
\hline
\end{tabular}}
\vspace{-2mm}
\caption{\label{EpinionsTable}Computed balance ratios $R_\ell$ for $1\leq\ell\leq 20$ on the Epinions network together with twice the standard deviation exhibited by the Monte Carlo results. We found no simple cycle of length 16, 18 or 20 on this network. Furthermore, we were unable to determine the standard deviation of the balance for $R_{17}$ and $R_{19}$. Regardless of the accuracy on these results, the small numbers of simple cycles of such lengths that we found imply that the null-hypothesis is compatible with all values  $15\%\lesssim R_{\ell}\lesssim 85\%$, as per Eq.~(\ref{ErrorNull}). 
}
\end{table*}
\begin{table*}[h!]
\renewcommand{\arraystretch}{1.1}
\resizebox{\textwidth}{!}{
\begin{tabular}{|c|c|c|c|c|c|}
\hline
\hline  
&\textbf{Self-loops $\mathcal{U}_1$}& \textbf{Backtracks $\mathcal{U}_2$} & \textbf{Triangles $\mathcal{U}_3$}& \textbf{Squares $\mathcal{U}_4$}& \textbf{Pentagons $\mathcal{U}_5$}\\
\hline
\textbf{Monte Carlo}  & $6.5056\pm0.0003\%$& $2.1301\pm0.0004\%$ & $12.658\pm0.0028\%$ &$ 16.51\pm0.004\%$ &$18.50\pm0.014\%$\\
\hline 
 \textbf{Exact}                  & $6.5056\%$ & $2.1302\%$ & $12.656\%$ &  & \\
 \hline 
 \multicolumn{6}{c}{}\\
 \hline
& \textbf{ $\mathcal{U}_6$}&  \textbf{ $\mathcal{U}_7$}& \textbf{ $\mathcal{U}_8$}& \textbf{ $\mathcal{U}_9$}& \textbf{ $\mathcal{U}_{10}$}\\
\hline
\textbf{Monte Carlo} & $19.15\pm0.04\%$ &$19.03\pm0.24\%$ &$19.01\pm1.16\%$& $20.96\pm3.27\%$&$28.09\pm5.58\%$\\ 
 \hline 
 \multicolumn{6}{c}{}\\
 \hline
& \textbf{ $\mathcal{U}_{11}$}&  \textbf{ $\mathcal{U}_{12}$}& \textbf{ $\mathcal{U}_{13}$}& \textbf{ $\mathcal{U}_{14}$}& \textbf{ $\mathcal{U}_{15}$}\\
\hline
\textbf{Monte Carlo} & $78.2 \pm 10.3\%$ &$92.7\pm2.6\%$ &$112.5\pm7.7\%$&$164.6\pm4.0\%$&$90.9\pm2.1\%$\\ 
 \hline 
 \multicolumn{6}{c}{}\\
 \hline
& \textbf{ $\mathcal{U}_{16}$}&  \textbf{ $\mathcal{U}_{17}$}& \textbf{ $\mathcal{U}_{18}$}& \textbf{ $\mathcal{U}_{19}$}& \textbf{ $\mathcal{U}_{20}$}\\
\hline
\textbf{Monte Carlo} &  &$100\%$ && $79.9\%$&\\ 
\hline
\hline
\end{tabular}}
\vspace{-2mm}
\caption{\label{EpinionsTable2}Computed ratio of negative to positive simple cycles $\mathcal{U}_\ell$ for $1\leq\ell\leq 20$ on the Epinions network together with twice the standard deviation exhibited by the Monte Carlo results.}
\end{table*}
\begin{table*}[h!]
\renewcommand{\arraystretch}{1.1}
\resizebox{\textwidth}{!}{
\begin{tabular}{|c|c|c|c|c|c|}
\hline
\hline  
&\textbf{Self-loops $K_1$}& \textbf{Backtracks $K_2$} & \textbf{Triangles $K_3$}& \textbf{Squares $K_4$}& \textbf{Pentagons $K_5$}\\
\hline
\textbf{Monte Carlo}  & $0.87784\pm0.00001$& $0.95829\pm0.00001$ & $0.77529\pm0.00005$ &$0.7166\pm0.0001$ &$0.6878\pm0.0002$\\
\hline 
 \textbf{Exact}                  & $0.87784$ & $0.95828$ & $0.77531$ &  & \\
 \hline 
 \multicolumn{6}{c}{}\\
 \hline
& \textbf{ $K_6$}&  \textbf{ $K_7$}& \textbf{ $K_8$}& \textbf{ $K_9$}& \textbf{ $K_{10}$}\\
\hline
\textbf{Monte Carlo} & $0.679\pm0.001$ &$0.68\pm0.003$ &$0.681\pm0.016$& $0.654\pm0.045$&$0.563\pm0.08$\\ 
 \hline 
 \multicolumn{6}{c}{}\\
 \hline
& \textbf{ $K_{11}$}&  \textbf{ $K_{12}$}& \textbf{ $K_{13}$}& \textbf{ $K_{14}$}& \textbf{ $K_{15}$}\\
\hline
\textbf{Monte Carlo} & $0.124\pm0.065$ &$0.038\pm0.014$ &$-0.058\pm0.034$&$-0.244\pm0.011$&$0.048\pm0.011$\\ 
 \hline 
 \multicolumn{6}{c}{}\\
 \hline
& \textbf{ $K_{16}$}&  \textbf{ $K_{17}$}& \textbf{ $K_{18}$}& \textbf{ $K_{19}$}& \textbf{ $K_{20}$}\\
\hline
\textbf{Monte Carlo} &  &$0$ && $0.112$&\\ 
\hline
\hline
\end{tabular}}
\vspace{-2mm}
\caption{\label{EpinionsTable3}Computed relative signed cluster coefficient $K_\ell$ for $1\leq\ell\leq 20$ on the Epinions network together with twice the standard deviation exhibited by the Monte Carlo results.}
\end{table*}
\clearpage

\bibliographystyle{plain}

\begin{thebibliography}{00}

\bibitem{Alon1997}
Alon, N., Yuster, R. {\&} Zwick, U. (1997)  Finding and counting given length
  cycles. {\em Algorithmica}, \textbf{17}, 209--223.

\bibitem{Antal2006}
Antal, T., Krapivsky, P.~L. {\&} Red, S. (2006)  Social balance on networks:
  The dynamics of friendship and enmity. {\em Physica D: Nonlinear Phenomena},
  \textbf{224}, 130--136.

\bibitem{Aref2015}
Aref, S. {\&} Wilson, M.~C. (2015)  Measuring Partial Balance in Signed
  Networks. {\em arXiv:1509.04037}.

\bibitem{Avis1996}
Avis, D. {\&} Fukuda, K. (1996)  {Reverse search for enumeration}. {\em
  Discrete Applied Mathematics}, \textbf{65}, 21--46.

\bibitem{bax1996finite}
Bax, E. {\&} Franklin, J. (1996)  A finite-difference sieve to count paths and
  cycles by length. {\em Information Processing Letters}, \textbf{60}(4),
  171--176.

\bibitem{Birmele2013}
Birmel\'e, E., Ferreira, R., Grossi, R., Marino, A., Pisanti, N., Rizzi, R.
  {\&} Sacomoto, G. (2013)  {Optimal listing of cycles and st-paths in
  undirected paths}. In {\em Proceedings of the Twenty-Fourth Annual ACM-SIAM
  Symposium on Discrete Algorithms}, pages 1884--1896.

\bibitem{Borgwardt2005}
Borgwardt, K.~M. {\&} Kriegel, H.-P. (2005)  {Shortest-path kernels on graphs}.
  In {\em Proceedings of the 5th IEEE International Con- ference on Data Mining
  (ICDM 2005), 27-30 November 2005}.

\bibitem{Burgisser1997}
B\"{u}rgisser, P., Clausen, M. {\&} Shokrollahi, M.~A. (1997) {\em {Algebraic
  Complexity Theory}}, volume 315 of {\em Grundlehren der mathematischen
  Wissenschaften}.
Springer, Berlin; New York.

\bibitem{Cartwright1956}
Cartwright, D. {\&} Harary, F. (1956)  Structural balance: a generalization of
  Heider's theory. {\em Psychological Review}, \textbf{63}, 277--293.

\bibitem{Chiang2014}
Chiang, K.-Y., Hsieh, C.-J., Natarajan, N. {\&} Dhillon, I.~S. (2014)
  {Prediction and Clustering in Signed Networks: A Local to Global
  Perspective}. {\em Journal of Machine Learning Research}, \textbf{15},
  1177--1213.

\bibitem{Elbassioni2015}
Elbassioni, K. (2015)  {A Polynomial Delay Algorithm for Generating Connected
  Induced Subgraphs of a Given Cardinality}. {\em Journal of Graph Algorithms
  and Applications}, \textbf{19}, 273--280.

\bibitem{StanfordEpinions}
Epinions social network (2016)  {Epinions social network}.
  \url{http://snap.stanford.edu/data/soc-sign-epinions.html}.

\bibitem{Estrada2014}
Estrada, E. {\&} Benzi, M. (2014)  Walk-based measure of balance in signed
  networks: Detecting lack of balance in social networks. {\em Physical Review
  E}, \textbf{90}, 042802.

\bibitem{Facchetti2011}
Facchetti, G., Iacono, G. {\&} Altafini, C. (2011)  Computing global structural
  balance in large-scale signed social networks. {\em Proceedings of the
  National Academy of Sciences}, \textbf{108}, 20953--20958.

\bibitem{Giscard2016}
Giscard, P.-L. {\&} Rochet, P. (2016a)  Algebraic Combinatorics on Trace
  Monoids: Extending Number Theory to Walks on Graphs. {\em arXiv:1601.01780}.

\bibitem{Giscard2016b}
Giscard, P.-L. {\&} Rochet, P. (2016b)  Enumerating simple paths from connected
  induced subgraphs. {\em arXiv:1606.00289}.

\bibitem{Hage1973}
Hage, P. (1973)  A graph theoretic approach to the analysis of alliance
  structure and local grouping in highland New Guinea. {\em Anthropological
  Forum: A Journal of Social Anthropology and Comparative Sociology},
  \textbf{3}, 280--294.

\bibitem{Harary1953}
Harary, F. (1953)  On the notion of balance of a signed graph. {\em Michigan
  Math. J.}, \textbf{2}, 143--146.

\bibitem{Harary1959}
Harary, F. (1959)  On the measurement of structural balance. {\em Behavioral
  Science}, \textbf{4}, 316--323.

\bibitem{Harary1960}
Harary, F. (1960)  A matrix criterion for structural balance. {\em Naval
  Research Logistics}, \textbf{7}, 195--199.

\bibitem{Harary1980}
Harary, F. {\&} Kabell, J.~A. (1980)  A simple algorithm to detect balance in
  signed graphs. {\em Mathematical Social Sciences}, \textbf{1}, 131--136.

\bibitem{Harary1971}
Harary, F. {\&} Manvel, B. (1971)  On the number of cycles in a graph. {\em
  Matematick\'y \v{c}asopis}, \textbf{21}, 55--63.

\bibitem{Hashimoto1989}
Hashimoto, K. (1989)  {Zeta functions of finite graphs and representations of
  p-adic groups}. {\em Stud. Pure Math.}, \textbf{15}, 211--280.

\bibitem{Heider1946}
Heider, F. (1946)  Attitudes and cognitive organization. {\em The Journal of
  Psychology}, \textbf{21}, 107--112.

\bibitem{Horvath2004}
Horv\'{a}th, T., Gartner, T. {\&} Wrobel, S. (2004)  {Cyclic pattern kernels
  for predictive graph mining}. In {\em Proceedings of the Tenth ACM SIGKDD
  International Conference on Knowledge Discovery and Data Mining}.

\bibitem{Iacono2010}
Iacono, G., Ramezani, F., Soranzo, N. {\&} Altafini, C. (2010)  {Determining the
  distance to monotonicity of a biological network: a graph-theoretical
  approach}. {\em IET Systems Biology}, \textbf{4}, 223--235.

\bibitem{Ihara1966}
Ihara, Y. (1966)  On discrete subgroups of the two by two projective linear
  group over p-adic fields. {\em Journal of the Mathematical Society of Japan},
  \textbf{18}, 219--235.

\bibitem{Impagliazzo2001}
Impagliazzo, R. {\&} Paturi, R. (2001)  {On the Complexity of $k$-SAT}. {\em
  Journal of Computer and System Sciences}, \textbf{62}, 367--375.

\bibitem{Johnson1975}
Johnson, D.~B. (1975)  {Finding all the elementary circuits of a directed
  graph}. {\em SIAM J. Comput.}, \textbf{4}, 77--84.

\bibitem{Kunegis2014}
Kunegis, J. (2014)  {Applications of Structural Balance in Signed Social
  Networks}. {\em arXiv:1402.6865}.

\bibitem{Kunegis2010}
Kunegis, J., Schmidt, S., Lommatzsch, A.~S., Lerner, J., DeLuca, E.~W. {\&}
  Albayr, S. (2010)  {Spectral analysis of signed graphs for clustering,
  prediction and visualization}. {\em SIAM Conference on Data Mining (Society
  for Industrial and Applied Mathematics, Philadelphia)}, pages 559--570.

\bibitem{Lampe2007}
Lampe, C.~A., Johnston, E. {\&} Resnick, P. (2007)  {Follow the reader:
  Filtering comments on Slashdot}. {\em Proceedings of Computer/Human
  Interaction 2007 Conference (Association for Computing Machinery, New York)},
  pages 1253--1262.

\bibitem{Lange2015}
Lange, C., Liu, S., Peyerimhoff, N. {\&} Post, O. (2015)  {Frustration index
  and Cheeger inequalities for discrete and continuous magnetic Laplacians}.
  {\em Calculus of variations and partial differential equations}, \textbf{54},
  4165--4196.

\bibitem{Leskovec2010}
Leskovec, J., Huttenlocher, D. {\&} Kleinberg, J. (2010)  {Signed networks in
  social media}. {\em Conference on Human Factors in Computing Systems}.

\bibitem{Mateti1976}
Mateti, P. {\&} Deo, N. (1976)  {On Algorithms for Enumerating all Circuits of
  a Graph}. {\em SIAM J. Comput.}, \textbf{5}, 90--99.

\bibitem{Movarraei2016}
Movarraei, N. {\&} Boxwala, S.~A. (2016)  {On the Number of Cycles in a Graph}.
  {\em Open Journal of Discrete Mathematics}, pages 41--69.

\bibitem{Norman1972}
Norman, R.~Z. (1972)  A derivation of a measure of relative balance for social
  structures and a characterization of extensive ratio systems. {\em Journal of
  Mathematical Psychology}, \textbf{9}, 66--91.

\bibitem{Giscard2016c}
P.-L.~Giscard, N.~Kriege, R. C.~W. (2016)  {A general purpose algorithm for
  counting simple cycles and simple paths of any length}. {\em
  arXiv:1612.05531}.

\bibitem{Aref2016}
S.~Aref, A. J.~Mason, M. C.~W. (2016)  {An exact method for computing the
  frustration index in signed networks using binary programming}. {\em
  arXiv:1611.09030}.

\bibitem{StanfordSlashdot}
Slashdot social network, February 2009 (2016)  {Slashdot social network,
  February 2009}.
  \url{http://snap.stanford.edu/data/soc-sign-Slashdot090221.html}.

\bibitem{Stark1996}
Stark, H.~M. {\&} Terras, A.~A. (1996)  Zeta functions of finite graphs and
  coverings. {\em Advances in Mathematics}, \textbf{121}, 124--165.

\bibitem{Terras2011}
Terras, A.~A. (2011) {\em Zeta Functions of Graphs: A Stroll through the
  Garden}.
Cambridge University Press, Cambridge, 1st edition.

\bibitem{Terzi2011}
Terzi, E. {\&} Winkler, M. (2011)  {A spectral algorithm for computing social
  balance}. {\em Proceedings of the 8th international conference on Algorithms
  and models for the web graph}, pages 1--13.

\bibitem{Gama}
UCINET IV Datasets (2016)  {UCINET IV Datasets, Read Highland Tribes}.
  \url{http://vlado.fmf.uni-lj.si/pub/networks/data/ucinet/ucidata.htm}.

\bibitem{Uehara1999}
Uehara, R. (1999)  {The number of connected components in graphs and its
  applications}. {\em IEICE Technical Report, COMP99-10}.

\bibitem{StanfordWiki}
Wikipedia adminship election data (2016)  {Wikipedia adminship election data}.
  \url{http://snap.stanford.edu/data/wiki-Elec.html}.

\bibitem{Zajonc1965}
Zajonc, R.~B. {\&} Burnstein, E. (1965)  Structural balance, reciprocity, and
  positivity as sources of cognitive bias. {\em Journal of Personality},
  \textbf{33}, 570--583.

\end{thebibliography}

\end{document}